\documentclass[]{IEEEtran}%
\usepackage{color}
\usepackage{ifpdf}
\usepackage{multicol}
\usepackage{cite}
  \usepackage{graphicx}
\usepackage[cmex10]{amsmath}
\usepackage{algorithmic}
\usepackage{array}

\usepackage{mdwmath}
\usepackage{mdwtab}

\usepackage{eqparbox}
\usepackage[tight,normalsize,sf,SF]{subfigure}
\usepackage {url}
\usepackage {breqn}

\usepackage{epsfig}
\usepackage {amsfonts}
\usepackage[hang,small,bf]{caption}

\usepackage{rotating}
\usepackage{multirow}

\allowdisplaybreaks

\newcommand{\tabincell}[2]{\begin{tabular}[t]{@{}#1@{}}#2\end{tabular}}

\newcommand{ \blue }[1]{{\color{blue}#1}}

\usepackage[normalem]{ulem}
\newcommand{\off}[1]{}

\newcommand{\on}[1]{#1}

\begin{document}

%
\title{A Survey on Device-to-Device Communication in Cellular Networks}
\author{Arash Asadi,~\IEEEmembership{Student Member,~IEEE,}
Qing Wang,~\IEEEmembership{Student Member,~IEEE,}
        and~Vincenzo Mancuso,~\IEEEmembership{Member,~IEEE}
\IEEEcompsocitemizethanks{\IEEEcompsocthanksitem ~~~\protect\\
\IEEEcompsocthanksitem  The authors are with the Institute IMDEA Networks, Madrid, Spain and University Carlos III of Madrid, Spain (email: \{arash.asadi, qing.wang, vincenzo.mancuso\}@imdea.org).}
\thanks{}}



\IEEEcompsoctitleabstractindextext{%

\begin{abstract}
Device-to-Device (D2D) communication was initially proposed in cellular networks as a new paradigm to enhance network performance. 
The emergence of new applications such as content distribution and location-aware advertisement introduced new use-cases for D2D 
communications in cellular networks. The \on{initial} studies show\on{ed} that D2D communication has advantages such as increased spectral efficiency and reduced communication delay. However, this communication mode introduces complications in terms of interference control overhead and protocols that are still open research problems.
\off{Academia and industry have been trying to put D2D communications into work for the past decade. In fact,} The feasibility of D2D communications in LTE-A is being studied by academia, industry, and the standardization bodies. To date, there are more than 100 papers available on D2D communications in cellular networks and, there is no survey on this field.\off{D2D communications in cellular networks.} In this article, we provide a taxonomy based on the D2D communicating spectrum and review the available literature extensively under the proposed taxonomy. Moreover, we provide new insights into the over-explored and under-explored areas which lead us to identify open research problems of D2D communication in cellular networks.   
\end{abstract}

}

\maketitle

\IEEEdisplaynotcompsoctitleabstractindextext
\IEEEpeerreviewmaketitle

\section{Introduction}
As telecom operators are struggling to accommodate the existing demand of mobile users, new data \off{consuming} \on{intensive} applications are emerging in daily routines of mobile users (e.g., proximity-aware services). Moreover, 4G cellular technologies (WiMAX~\mbox{\cite{wimax}} and LTE-A~\mbox{\cite{LTEadv36213}}), which have extremely efficient physical and MAC layer performance, are still lagging behind mobile users' booming data demand. Therefore, researchers are seeking for new paradigms to revolutionize the traditional communication methods of cellular networks. Device-to-Device (D2D) communication is one of such paradigms that appears to be a promising component in next generation cellular technologies. 

\begin{figure} [!t]
\centering
\includegraphics[scale=0.45]{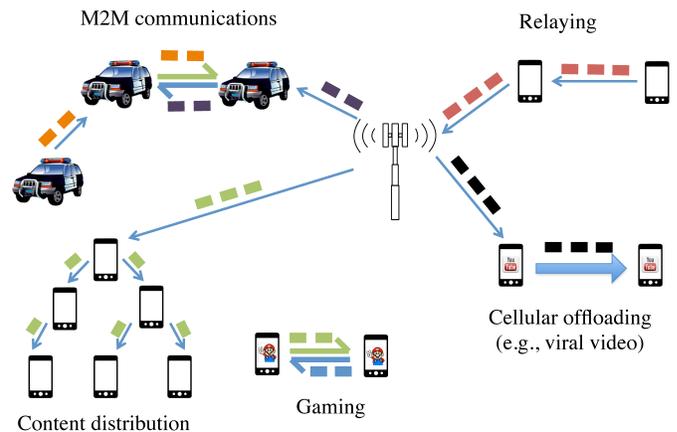}
\caption{Representative use-cases of D2D communications in cellular networks.}
\vspace{-2mm}
\label{fig:use-case}
\vspace{-4mm}
\end{figure}

\on{D2D communication in cellular networks is defined as direct communication between two mobile users without traversing the Base Station (BS) or core network. D2D communication is generally non-transparent to the cellular network and it can occur on cellular spectrum (i.e., {\it inband}) or unlicensed spectrum (i.e., {\it outband}). 
In a traditional cellular network, all communications must go through the BS even if both communicating parties are in range for D2D communication. This architecture suits the conventional low data rate mobile services such as voice call and text message in which users are not usually close enough to have direct communication. However, mobile users in today's cellular networks use high data rate services (e.g., video sharing, gaming, proximity-aware social networking) in which they could potentially be in range for direct communications (i.e., D2D). Hence, D2D communications in such scenarios can highly increase the spectral efficiency of the network. \on{Nevertheless, the advantages of D2D communications is not only limited to enhanced spectral efficiency.} In addition to improving spectral efficiency, D2D communications can potentially improve throughput, energy efficiency, delay, and fairness.}

\on{In academia, D2D communication was first proposed in~\cite{lin_multihop_2000} to enable multihop relays in cellular networks. Later the works in~\cite{kaufman_cellular_2008,doppler_device_2009,doppler_device_2009A,osseiran_advances_2009A,peng_interference_2009} investigated the potential of D2D communications for improving spectral efficiency of cellular networks. Soon after, other potential D2D use-cases were introduced in the literature such as multicasting~\cite{du_compressed_2012,zhou_intracluster_2013}, peer-to-peer communication~\cite{lei_operator_2012}, video dissemination~\cite{golrezaei_base_2012,golrezaei2012device,li_device_2012,doppler_device_2009}, machine-to-machine (M2M) communication~\cite{pratas_low_2013}, cellular offloading~\cite{bao_dataspotting_2010}, and so on. The most popular use-cases of D2D communications are shown in Fig.~\ref{fig:use-case}\on{.}}
\on{The first attempt to implementing D2D communication \off{capability} in a cellular network was made by Qualcomm's FlashLinQ~\cite{wu2010Allert} which is a PHY/MAC network architecture for D2D communications underlaying cellular networks. FlashLinQ takes advantage of OFDM/OFDMA technologies and distributed scheduling to create an efficient method for timing synchronization, peer discovery, and link management in D2D-enabled cellular networks.} \off{A 3GPP study group is investigating D2D communications in cellular networks} 
\on{ In addition to academia and telecommunication companies, 3GPP is also investigating D2D communications as Proximity Services (ProSe). In particular, the feasibility of ProSe and its use-cases in LTE are studied in~\cite{3GPPTR22.803} and the required architectural enhancements to accommodate such use-cases are investigated in~\cite{3GPPTR23.703}. Currently, ProSe is supposed to be included in 3GPP Release $12$ as a public safety network feature with focus on one to many communications~\cite{3GPPTR23.703}. A brief overview of standardization activities and fundamentals of 3GPP ProSe can be found in~\cite{Lin2013ComMag}}.

\begin{figure} [!t]
\centering
\includegraphics[scale=0.42]{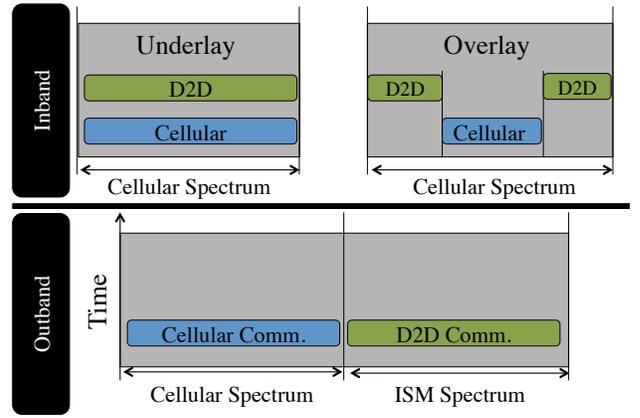}
\caption{Schematic representation of overlay inband, underlay inband, and outband D2D.}
\vspace{-2mm}
\label{fig:in-outband}
\vspace{-4mm}
\end{figure}

The majority of the literature \off{in} \on{on} D2D communications propose to use the cellular spectrum for both D2D and cellular communications (i.e., {\it underlay inband} D2D). These works usually study the problem of interference mitigation between D2D and cellular communication~\cite{yu_performance_2009,xu_interference_2012,xu_effective_2010,xu_performance_2012,peng_interference_2009,zhang_interference_2013,janis_interference_2009,min_capacity_2011,elkotby_exploiting_2012}.
In order to avoid the aforementioned interference issue, some propose to dedicate part of the cellular resources only to D2D communications (i.e., {\it overlay inband} D2D).
Here resource allocation gains utmost importance so that dedicated cellular resources be not wasted~\cite{pei_resource_2013}.
Other researchers propose to adopt outband rather than inband D2D communications in cellular networks so that the precious cellular spectrum be not affected by D2D communications. In outband communications, the coordination between radio interfaces is either controlled by the BS (i.e, {\it controlled}) or the users themselves (i.e., {\it autonomous}).
Outband D2D communication faces a few challenges in coordinating the communication over two different bands because usually D2D communication happens on a second radio interface (e.g., WiFi Direct~\cite{alliance2010wi} and bluetooth~\cite{bluetooth2001bluetooth}). The studies on outband D2D investigate issues such as power consumption~\cite{asadi2013energy,arash2013MSWIM, arash2013wireless, wang2013WOWMOM} and inter-technology architectural design. \on{Fig.~\ref{fig:in-outband} graphically depicts the difference among underlay inband, overlay inband, and outband communications. }
\subsection{Related Topics}
\on{Since D2D communication is a new trending topic in cellular networks, there is no survey available on the topic. However, from an architectural perspective, D2D communications may look similar to Mobile Ad-hoc NETworks (MANET) and Cognitive Radio Networks (CRN). However, there are some key differences among these architectures that can not be ignored. Although there is no standard for D2D communications, D2D communications in cellular network are expected to be overseen/controlled by a central entity (e.g., evolved Node B (eNB)). D2D users may act autonomously only when the cellular infrastructure is unavailable. The involvement of the cellular network in the control plane is the key difference between D2D, and MANET and CRN. The availability of a supervising/managing central entity in D2D communications resolves many existing challenges of MANET and CRN such as white space detection, collision avoidance, and synchronization. Moreover, D2D communication is mainly used for single hop communications, thus, it does not inherit the multihop routing problem of MANET. An extensive survey on spectrum sensing algorithms for cognitive radio applications and routing protocols for MANET can be found in~\cite{yucek2009ComSurvey} and~\cite{kumar2009SoftComp}, respectively.
M2M communication~\cite{3GPPM2M} is another architecture that might benefit from D2D-like schemes. M2M is the data communication between machines that does not necessarily need human interaction. Although M2M, similarly to D2D, focuses on data exchange between (numerous) nodes or between nodes and infrastructure, it does not have any requirements on the distances between the nodes. So, M2M is application-oriented and technology-independent while D2D aims at proximity connectivity services and it is technology-dependent.
}

\begin{figure*} [!t]
\centering
\includegraphics[scale=0.65]{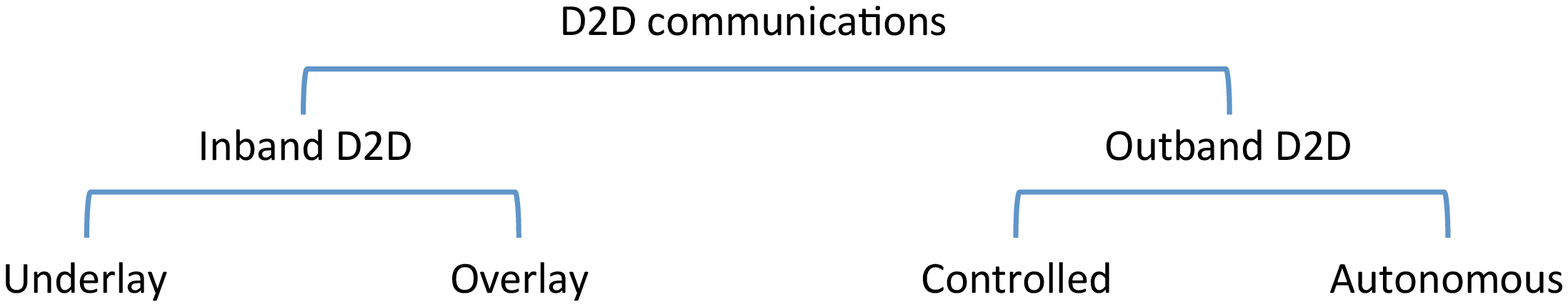}
\caption{Device-to-Device communication classification.}
\vspace{-2mm}
\label{fig:taxo}
\vspace{-4mm}
\end{figure*}

\subsection{Contributions and Organization of the Survey}
In this paper, we provide an extensive review of available literature on D2D communications, \on{which is the first of its kind.} Moreover, we provide new insights to the existing works which lead us to the \on{under-explored} open issues \off{that are not explored yet}. In Section~\ref{s:taxo}, we categorize the available literature based on our proposed taxonomy. In Sections~\ref{s:underlay}~and~\ref{s:overlay}, we review the works using inband D2D. The papers proposing to use outband D2D are surveyed in~Section~\ref{s:outband}. After reviewing the available literature, we discuss the state-of-the-art protocol proposals and provide an overview of 3GPP ProSe services in Section~\ref{s:protocols}. In Section~\ref{s:discussions}, we provide a discussion on the common assumptions of the surveyed literature, the advantages and disadvantages of different approaches, the maturity of the field, and its emergence into the real world systems. In addition, this section sheds light on the open issues and potential research directions in D2D communications. Finally, we conclude the paper in Section~\ref{s:conclusion}.

\section{Taxonomy}
\label{s:taxo}
In this section, we categorize the available literature on D2D communication in cellular networks based on the spectrum in which D2D communication occurs. In the following subsection we provide a formal definition for each category and sub-category. Next, we provide a quick overview to advantages and disadvantages of each D2D method.

{\bf Inband D2D:} The literature under this category, which contains the majority of the available work,  proposes to use the cellular spectrum for both D2D and cellular links. The motivation for choosing inband communication is usually the high control over cellular (i.e., licensed) spectrum. Some researchers (see e.g.,~\cite{akkarajitsakul_mode_2012, doppler_device_2009A}) \off{believe} \on{consider} that the interference in the unlicensed spectrum is uncontrollable which imposes constraints for QoS provisioning. Inband communication can be further divided into underlay and overlay categories. In underlay D2D communication, cellular and D2D communications share the same radio resources. In contrast, D2D links in overlay communication are given dedicated cellular resources. 
{Inband D2D can improve the spectrum efficiency of cellular networks by reusing spectrum resources (i.e., underlay) or allocating dedicated cellular resources to D2D users that accommodates direct connection between the transmitter and the receiver (i.e., overlay). The key disadvantage of inband D2D is the interference caused by D2D users to cellular communications and vice versa. This interference can be mitigated by introducing high complexity resource allocation methods, which increase the computational overhead of the BS or D2D users.}

{\bf Outband D2D:} Here the D2D links exploit unlicensed spectrum. The motivation behind using outband D2D communication is to eliminate the interference issue between D2D and cellular link. Using unlicensed spectrum requires an extra interface and usually adopts other wireless technologies such as WiFi Direct~\cite{alliance2010wi}, ZigBee~\cite{alliance2006zigbee} or bluetooth~\cite{bluetooth2001bluetooth}. Some of the work on outband D2D (see e.g.,~\cite{golrezaei_base_2012, golrezaei2012device, arash2013MSWIM, asadi2013energy}) suggest to give the control of the second interface/technology to the cellular network (i.e., controlled). In contrast, others (see e.g.,~\cite{wang2013WOWMOM}) propose to keep cellular communications controlled and leave the D2D communications to the users (i.e., autonomous). 
\off{Since outband D2D communications do not occur on cellular spectrum, there is no interference issue like in inband D2D.} \on{Outband D2D uses unlicensed spectrum which makes the interference issue between D2D and cellular users irrelevant. On the other hand, outband D2D may suffer from the uncontrolled nature of unlicensed spectrum.} It should be noted that only cellular devices with two wireless interfaces (e.g., LTE and WiFi) can use outband D2D, and thus users can have simultaneous D2D and cellular communications.

Fig.~\ref{fig:taxo} illustrates the taxonomy introduced for D2D communications in cellular networks. In the following sections, we review the related literature based on this taxonomy. 
\section{Underlaying Inband D2D}
\label{s:underlay}
Early works on D2D in cellular networks propose to reuse cellular spectrum for D2D communications. To date, the majority of available literature is also dedicated to inband D2D, especially D2D communications underlaying cellular networks. In this section, we review the papers that employ underlaying D2D to improve the performance of cellular networks, in terms of \emph{spectrum efficiency, energy efficiency, cellular coverage}, and other performance targets.

\subsection{Spectrum Efficiency}
By exploiting the spatial diversity, underlaying inband D2D is able to increase the cellular spectrum efficiency. This can be done by proper interference management, mode selection,\footnote{In general, mode selection involves choosing between cellular mode (i.e., the BS is used as a relay) and D2D mode (i.e., the traffic is directly transmitted  to the receiver).} resource allocation and by using network coding.

Interference between the cellular and D2D communications is the most important issue in underlaying D2D communications. Good interference management algorithms can increase the system capacity, and \off{has} \on{have} attracted a lot of attention~\cite{peng_interference_2009,xu_effective_2010, janis_interference_2009, kaufman_cellular_2008, min_capacity_2011,chen_downlink_2012, yu_device_2012}.
The authors of~\cite{kaufman_cellular_2008} propose to use cellular uplink resources for D2D communications. Since reusing uplink resources for D2D users can cause interference to cellular uplink transmissions at the BS, D2D users monitor the received power of downlink control signals to estimate the pathloss between D2D transmitter and the BS. This helps \on{the} D2D users to maintain the transmission power below a threshold to avoid high interference to cellular users. If the required transmission power for a D2D link is higher than the minimal interference threshold, the D2D transmission is not allowed. The authors also propose to use \off{Dynamic Source Routing} \on{dynamic source routing}~\cite{johnson1996dynamic} algorithm for routing among D2D users in case of multi-hop communications. The simulations show that probability of having D2D links increases with stronger pathloss component. This is because the stronger the pathloss, the weaker the interference caused by D2D transmission at the BS. In~\cite{peng_interference_2009}, the authors also study the uplink interference between D2D and cellular users and propose two mechanisms to avoid interference from cellular users to D2D users and vice versa. In order to reduce the interference from cellular users to D2D communications, D2D users read the resource block allocation information from \on{the} control channel. Therefore, they  can avoid using resource blocks that are used by the cellular users \on{in the proximity} \off{which are located close to them}. The authors propose to broadcast the expected interference from D2D communication on cellular resource block to all D2D users. Hence, the D2D users can adjust their transmission power and resource block selection in a manner that the interference from D2D communication to uplink transmission is below the tolerable threshold. The authors show via simulation that the proposed mechanisms improve the system throughput by $41\%$.\footnote{\on{Note that the numerical performance gains reported in this article may have been obtained under different simulation/experiment settings which are specific to the cited work.  }} 

Zhang {\it et al.}~\cite{zhang_interference_2013} propose a graph-based resource allocation method for cellular networks with underlay D2D communications. They mathematically formulate the optimal resource allocation as a non-linear problem which is NP-Hard. The authors propose a suboptimal graph-based approach which accounts for interference and capacity of the network. In their proposed graph, each vertex represents a link (D2D or cellular) and each edge connecting two vertices shows the potential interference between the two links. The simulation results show that the graph-based approach performs close to the throughput-optimal resource allocation.

In~\cite{xu_effective_2010}, a new interference cancellation scheme is designed based on the location of users. The authors propose to allocate a dedicated control channel for D2D users. Cellular users listen to this channel and measure the SINR. If the SINR is higher than a pre-defined threshold, a report is sent to the eNB. Accordingly, the eNB stops scheduling cellular users on the resource blocks that are currently occupied  by D2D users. The eNB also sends broadcast information regarding the location of the users and their allocated resource blocks. Hence, D2D users can avoid using resource blocks which interfere with cellular users. Simulation results show that the interference cancellation scheme can increase the average system throughput up to $374\%$ in comparison to the scenario with no interference cancellation. 
Janis {\it et al.} address a similar solution in~\cite{janis_interference_2009}, where the D2D users also measure the signal power of cellular users and inform the BS of these values. The BS then avoids allocating the same frequency-time slot to the cellular and D2D users which have strong interference \off{on} \on{with} each other, which is different from~\cite{xu_effective_2010}. The proposed scheme of~\cite{janis_interference_2009} minimizes the maximum received power at D2D pairs from cellular users. The authors first show via numerical results that D2D communications with random resource allocation can increase the mean cell capacity over a conventional cellular system by $230\%$. Next, they show that their proposed interference-aware resource allocation scheme achieves $30\%$ higher capacity gain than the {random resource allocation strategy}.

\off{Unlike conventional D2D interference management mechanisms in which the interference is controlled by limiting D2D transmission power, the work in~\mbox{\cite{min_capacity_2011}} proposes a new interference management, i.e., use physical separation of cellular and D2D users for interference management.}
\on{The work in~\cite{min_capacity_2011} proposes a new interference management in which the interference is not controlled by limiting D2D transmission power as in the conventional D2D interference management mechanisms.} The proposed scheme defines an interference limited area in which no cellular users can occupy the same resources as the D2D pair. Therefore, the interference between the D2D pair and cellular users is avoided. The disadvantage of this approach is reducing multi-user diversity because the physical separation limits the scheduling alternatives for the BS. However, numerical simulations prove that the capacity loss due to multi-user diversity reduction is negligible compared to the gain achieved by their proposal. In fact, this proposal provides a gain of $129\%$ over conventional interference management schemes.
\on{A} similar method is also considered in~\cite{chen_downlink_2012}, where interference limited areas are formed according to the amount of tolerable interference and minimum SINR requirements for \on{successful} transmission. The proposed scheme consists in: $(i)$ defining interference limited areas where cellular and D2D users cannot use the same resource; and $(ii)$ 
{allocating the resources in a manner that D2D and cellular users within the same interference area use different resources.}
The simulation results show that the proposed scheme performs almost as good as Max-Rate~\cite{knopp95ICC} and better than conventional D2D schemes.

Yu {\it et al.}~\cite{yu_device_2012} propose to use Han-Kobayashi rate splitting techniques~\cite{han1981new} to improve the throughput of D2D communications. In rate splitting, the message is divided into two parts, namely, private and public. The private part, as the name suggests, can be decoded only by the intended receiver, and the public part can be decoded by any receiver. This technique helps D2D interference victims to cancel the interference from the public part of the message by running a best-effort successive interference cancellation algorithm~\cite{rasmussen2000matrix}. The authors also analytically solve the rate-splitting problem in a scenario with two interfering links. Finally, they show via numerical simulations that their rate splitting proposal increases the cell throughput up to \off{$6.5$ times} \on{$650\%$} higher when the D2D pair is placed far from the BS and close to each other.

Doppler {\it et al.} study different aspects of D2D communications in cellular networks in~\cite{doppler_device_2009,doppler_device_2009A,doppler_mode_2010,doppler_advances_2011}. They study the session and interference management in D2D communications as an underlay to LTE-A networks in~\cite{doppler_device_2009}. In this work, they mainly discuss the concepts of D2D and provide a first order protocol for the necessary functionality and signaling. They use numerical simulations to show that D2D enabled cellular networks can achieve up to $65\%$ higher throughput than conventional cellular networks.
In~\cite{doppler_mode_2010}, they study the problem of mode selection (i.e., cellular or D2D) in LTE-A cellular networks. They propose to estimate the achievable transmission rate in each mode by utilizing the channel measurements performed by users. After the rate estimation, each user chooses the mode which results in higher transmission rate at each scheduling epoch. The simulations show that their proposal has  $50\%$ gain on system throughput over the conventional cellular communications.

To improve the capacity of cellular networks, the authors of~\cite{osseiran_advances_2009} propose a joint D2D communication and network coding scheme. They consider \off{the} cooperative networks~\cite{sendonaris2003userI,sendonaris2003userII}, where  D2D communication is used to exchange uplink messages among cellular users before the messages are transmitted to the BS. For example, cellular users $a$ and $b$ exchange their uplink data over D2D link. Then each user sends the coded data containing the original data from both users to the BS. Here, the interference 
is controlled using the interference-aware algorithm proposed in~\cite{janis_interference_2009}.
They show that random selection of cooperative users is not efficient because the combination of users' channel qualities may not be suitable for network coding. To overcome this inefficiency, they propose to group the users with complementary characteristics to enhance the performance of network coding.
Using numerical simulation, they show that their proposal  increases the capacity by $34\%$ and $16\%$, in comparison to random selection and decode-and-forward relaying schemes, respectively. Moreover, they show that multi-antenna capability reduces the impact of interference from the BS and increases the number of D2D users by $30\%$.

\begin{figure}[!ht]
  \centering
  \includegraphics[width=80mm]{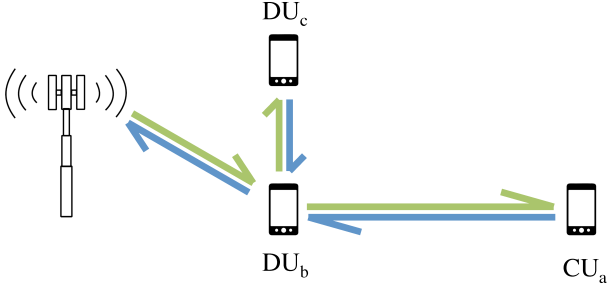}
  \caption{\on{The evaluation scenario of~\cite{pei_resource_2013}. The cell includes a D2D pair (i.e., $\text{DU}_b$ and $\text{DU}_c$) and one cellular user (i.e., $\text{CU}_a$). $\text{DU}_b$ also acts as a relay between the BS and $\text{CU}_a$.}}
  \label{fig:d2d-scenario}
\end{figure}

The authors of \cite{pei_resource_2013} consider a single cell scenario including a cellular user ($\text{CU}_a$) and a D2D pair ($\text{DU}_b$ and $\text{DU}_c$). $\text{DU}_b$ and $\text{DU}_c$ communicate with each other over the D2D link and $\text{CU}_a$ communicates with the BS by using $\text{DU}_b$ as a relay (see Fig.~\ref{fig:d2d-scenario}). The relay (i.e., $\text{DU}_b$) can communicate bi-directionally with the other D2D user $\text{DU}_c$, as well as assisting the transmission between the BS and the cellular user $\text{CU}_a$. The time is divided into two different periods: $(i)$ during the first period, $\text{DU}_c$ and either the BS or $\text{CU}_a$ send data to $\text{DU}_b$ concurrently; and $(ii)$ during the second period, $\text{DU}_b$ sends data to $\text{DU}_c$ and either the BS or $\text{CU}_a$. The authors investigate the achievable capacity region of the D2D and the cellular link. Simulation results show that by adjusting the power of BS and cellular device, the area of capacity region of the D2D link and BS-device link can be enlarged by up to $60\%$.

Xu \textit{et al.} in \cite{xu_resource_2012} consider the sum-rate optimization in a single cell scenario with underlayed D2D communications. \on{Using underlay D2D communication, the network can suffer from intra-cell interference.} \off{They assume the D2D and cellular communications use the same channel resource, which results in intra-cell interference and multiple D2D links exists.} They \off{formulate the problem as an} \on{adopt the} iterative combinatorial auction game \on{in their proposed spectrum resource allocation mechanism. In this game, spectrum resources are considered to be bidders that compete to obtain business and  D2D links are considered as goods or services that are waiting to be sold. The authors formulate the valuation of each resource unit for groups of D2D links. Based on this, they propose a non-monotonic descending price auction algorithm and show that the proposed algorithm can converge in a finite number of iterations. Moreover, the complexity of their proposal is lower than traditional combinatorial allocation schemes.} \off{investigate its properties}. In the simulation, the authors use WINNER II channel models~\cite{hentila2007matlab} and the simulation results show that the proposed scheme can improve the sum-rate up to 13\%, which varies with the number of spectrum resource units. 

\on{\textbf{Summary:} The surveyed literature in this subsection showed that D2D communication can improve the spectrum efficiency greatly. This improvement can be achieved by exploiting techniques such as interference reduction among cellular and D2D users \cite{kaufman_cellular_2008, peng_interference_2009, janis_interference_2009, yu_device_2012, doppler_mode_2010, xu_resource_2012, xu_effective_2010} or interference aware/avoidance~\cite{min_capacity_2011, chen_downlink_2012,zhang_interference_2013, osseiran_advances_2009}. Among these papers, \cite{osseiran_advances_2009, xu_resource_2012} and \cite{zhang_interference_2013} adopt more advanced mathematical techniques than the others. The proposed methods in these papers can be either self-organized \cite{kaufman_cellular_2008} or network controlled \cite{peng_interference_2009, janis_interference_2009, yu_device_2012, doppler_mode_2010, xu_effective_2010,min_capacity_2011, chen_downlink_2012,osseiran_advances_2009, xu_resource_2012, zhang_interference_2013}.
The self-organized methods proposed in~\cite{kaufman_cellular_2008} introduce less overhead and are more efficient in comparison to network-controlled methods. It should also be noted that using advanced mathematical techniques, such as non-linear programming~\cite{zhang_interference_2013} and game theory~\cite{xu_resource_2012}, can result in higher gain than simpler interference reduction/avoidance methods based on heuristics. However, they also introduce higher computational overhead which should be taken into account when comparing the performances of the proposals.
}

\subsection{Power Efficiency}
Power efficiency enhancement techniques for D2D-enabled cellular networks is also a very interesting research topic. Xiao {\it et al.}~\cite{xiao_qos_2011} propose a heuristic algorithm for power allocation in OFDMA-based cellular networks. 
They propose a heuristic that performs power allocation and mode selection using the existing subcarrier and bit allocation algorithms in~\cite{zhang2004subcarrier}~and~\cite{youjun2006qos}. The heuristic first allocates the resources for the cellular users and then performs resource allocation and mode selection for D2D users. If the required power level for D2D transmission is higher than a certain threshold, the D2D pair communicates through the BS. 
Via simulations, they show that the integration of their proposed heuristic with the existing algorithms in \cite{zhang2004subcarrier,youjun2006qos} improves the downlink power consumption of the network around $20\%$ in comparison to the traditional OFDMA system without D2D.
 
The authors of~\cite{jung_joint_2012} propose an algorithm for power allocation and mode selection in D2D communication underlaying cellular networks. The algorithm measures the power efficiency, which is a function of transmission rate and power consumption, of the users in different modes (cellular  and D2D). After computing the power efficiency, each device uses the mode in which it achieves higher power efficiency. The drawback of this algorithm is that the controller should perform exhaustive search for all possible combinations of modes for all devices. The authors benchmark their algorithm against the scheme of~\cite{hakola_device_2010} in which two users communicate over D2D link only if their pathloss is \off{less} \on{lower} than the pathlosses between each user and the BS. The simulation results indicate that their algorithm achieves up to $100\%$ gain over the scheme proposed in~\cite{hakola_device_2010}. 

The authors of~\cite{belleschi_performance_2011} aim to minimize the overall transmission power in a multi-cell OFDM cellular network. They assume a multi-cell scenario in which the BS serves a fixed number of cellular and D2D users. The authors formulate the problem of joint mode selection, resource allocation, and power allocation through linear programming which is proven to be NP-Hard in a strong sense~\cite{gary1979computers}. Due to the complexity of linear programming, the authors decide to  consider the power allocation in a single cell and propose a heuristic algorithm to solve it. They use a distributed sub-optimal heuristic which performs mode selection and resource allocation in a single cell scenario. The performance of the heuristic is compared with other two schemes: $(i)$ cellular mode in which transmission should go through the BS; and $(ii)$ D2D mode in which all D2D users can only communicate directly and passing through the BS is not allowed. 
The authors provide simulation results showing that the gain of power efficiency of the proposed method over conventional cellular networks is significant (up to 100\%) when the distance between D2D users is less than 150m. 

\on{\textbf{Summary:} It was observed in this subsection that D2D communication can result in increased power efficiency of the network. A common technique to achieve this is to dynamically switch between cellular and D2D modes. The authors in~\cite{xiao_qos_2011} and \cite{belleschi_performance_2011} propose heuristic algorithms to solve the mode selection problem, while~\cite{jung_joint_2012} employs the brute-force technique. The performance of the method in~\cite{jung_joint_2012} is thus better than those in the other two, but it also requires much more computations. 
}

\subsection{Performance with QoS/Power Constraints} 
There are many works which focus on the improving the system performance while maintaining certain QoS/power constraints~\cite{feng_device_2013,su_resource_2013,han_subchannel_2012,le_fair_2012, yu_power_2009, chia_resource_2011}. The authors of~\cite{feng_device_2013} propose a resource allocation method for D2D communication underlaying cellular network, which guarantees QoS requirements for both D2D and cellular users. They mathematically formulate the resource allocation problem, which is a non-linear constraint optimization problem. They divide the problem into three subproblems. First, the BS checks the feasibility of the D2D connection based on the SINR requirements (admission control). Next, they formulate the optimal power control for the D2D pair. Finally, a maximum-weight bipartite matching based scheme~\cite{cheng1996maximum} is used for resource allocation for cellular and D2D users. The authors benchmark their proposed algorithm against the works in~\cite{kaufman_cellular_2008,janis_interference_2009,zulhasnine_efficient_2010} via numerical simulations. The results show that their approach provides up to $70\%$ throughput gain over the algorithms proposed in~\cite{kaufman_cellular_2008, janis_interference_2009, zulhasnine_efficient_2010}.

The authors of \cite{su_resource_2013} consider the mode selection and resource allocation in D2D communications underlay cellular networks, where several pairs of D2D links co-exist with several cellular users. They formulate the problem of maximizing the system throughput with minimum data rate requirements, and use the \emph{particle swarm optimization}~\cite{kennedy2010particle} method to obtain the solutions. \off{Through simulation, the authors show that, compared to the orthogonal resource sharing scheme (i.e., the overlay case, which will be explained later), their proposed method can improve the system throughput by about $15\%$, where the achievable gain varies with the distance of D2D users.} 
\on{The simulation results show that the proposed method has $15\%$ throughput gain over the orthogonal resource sharing scheme (i.e., overlay D2D which will be explained later), where the achievable gain varies with the distance of D2D users.} Simulation results also show that this method can improve the system performance under the constraint of minimum data rate of users.

The authors of \cite{han_subchannel_2012} consider the scheduling and mode selection problem for D2D in OFDMA networks. They assume that the system time is slotted and each channel is divided into sub-channels. They formulate the problem of maximizing the mean sum-rate of the system with QoS satisfaction as a stochastic optimization problem, and use the stochastic sub-gradient algorithm to solve it. \on{From the solution, they design a sub-channel opportunistic scheduling algorithm that takes into account the CSI of D2D and cellular links as well as the QoS requirement of each D2D user.} The numerical results show that the mean sum-rate can be improved by up to $500\%$. \on{This gain increases when the average D2D pair distance reduces.} Moreover, with the D2D communication, the fairness among users can be achieved with the QoS requirement specified for each user.

In~\cite{le_fair_2012}, a two-phase resource allocation scheme for cellular network with underlaying D2D communications is proposed. The authors first formulate the optimal resource allocation policy as an integer programming problem~\cite{schrijver1998theory} which is NP-Hard. Hence, they propose a two-phase low-complexity suboptimal solution instead of the NP-Hard problem. In the first phase, they extend the technique used in~\cite{kim2006use} to perform optimal resource allocation for cellular users. In the second phase, they use a heuristic subchannel allocation scheme for D2D flows which initiates the resource allocation from the flow with minimum rate requirements. The heuristic also accounts for a D2D power budget (i.e., a D2D transmission power that does not impact the transmission rate of cellular flows) in subchannel allocation for D2D flow.

The authors of~\cite{yu_power_2009} and \cite{chia_resource_2011} consider a single cell scenario where a cellular user and two D2D users share the same radio resources. They assume that the BS is aware of the instantaneous Channel State Information (CSI) of all the links and it controls the transmit power and the radio resources of the D2D links. The objective is to optimize the sum-rate with energy/power constraint, under three different link sharing strategies, i.e., non-orthogonal sharing mode, orthogonal sharing mode, and cellular mode. The authors show analytically that an optimal solution can be given either in closed-form or can be chosen from a set. The numerical simulation for single-cell~\cite{yu_power_2009}~and~\cite{chia_resource_2011},  and multi-cell~\cite{chia_resource_2011} scenarios illustrates that under their proposed algorithm, the D2D transmission will not bring much interference to the cellular transmission. Moreover, the interference-aware resource allocation increases the system sum-rate by up to $45\%$. Similar scenario and objective are considered in~\cite{yu_performance_2009}. The difference among~\cite{yu_performance_2009, yu_power_2009} and \cite{chia_resource_2011} is that in~\cite{yu_performance_2009} the BS is only aware of the average CSI of the links,  whereas in~\cite{yu_power_2009} and \cite{chia_resource_2011} it is aware of the instantaneous CSI of links.

\on{\textbf{Summary:} Improving the performance of D2D-enabled cellular systems with QoS/power constraints usually requires advanced techniques such as  stochastic optimization, non-linear programming, and integer optimization. As expected, the solution of these approaches and their derived sub-optimal heuristic can indeed improve the system performance with QoS/power constraints. However, they do not seem to be good candidate for time-stringent application with limited computational capacity. Nonetheless, the authors of~\cite{yu_power_2009} and \cite{chia_resource_2011} derived the closed-form of the optimal solution, that in fact reduces the computational complexity. It should be noted that their considered scenario only consist of a cellular user and a D2D pair which is not practical in reality.
}

\subsection{Miscellaneous}
\off{Despite of} \on{In addition to}  spectrum efficiency, power efficiency, and system performance with different constraints, there are some other interesting works aiming to enhance spectrum utility~\cite{cheng_resource_2012}, cellular coverage~\cite{vanganuru_system_2012}, fairness~\cite{xu_interference_2012} and reliability~\cite{min_reliability_2011, kaufman2013ton}. 
The authors of \cite{cheng_resource_2012} aim to improve the user spectrum utility through mode selection and power allocation, where the spectrum utility is defined as the combination of users' data rates, power expenditure and bandwidth. \off{The mode of communication can dynamically switch between BS mode (i.e., BS acts as a relay) and D2D mode (i.e., devices communicate with each other directly).} \on{ As for the mode selection, the users can choose to transmit in BS or D2D modes. In BS mode, D2D transmitter and receiver communicate through the BS as in conventional cellular system. In D2D mode, D2D transmitter directly communicates with the receiver using the cellular resources as in underlay D2D communication.} The authors first derive the optimal transmission power for the above mentioned modes, and then use evolutionary game~\cite{weibull1997evolutionary} to obtain the model selection. Each user performs mode selection individually and independently. The BS collects users' mode selection decisions and broadcast\on{s} this information to all \off{the} users to help them for future mode selections. Numerical results show that, via the proposed method, the spectrum utility can be improved by up to $33\%$ and $500\%$, when compared to solely BS mode and D2D mode, respectively.

Xu \textit{et. al} in~\cite{xu_interference_2012} propose a resource allocation method based on sequential second price auction for D2D communications underlaying cellular networks. In a second price auction, the winner pays as much as the second highest bid. In the proposed auction, each resource block is put on auction and D2D pairs should bid for the resource blocks that they want to occupy. Therefore, each D2D pair\off{s} make\on{s} a bidding for every resource block and the bidding values are a function of achievable throughput of the bidding D2D pair on the auctioned resource block. Simulation results show that the achievable throughput of their proposal is at least $80\%$ of the optimal resource allocation strategy. The results also illustrate that the proposal achieves a fairness index around $0.8$ and system sum-rate efficiency higher than $85\%$.

D2D communication is also a promising way to enlarge the cellular coverage and improve the performance of cell edge users, e.g., the authors of \cite{vanganuru_system_2012} propose a method to use nodes as virtual infrastructure to improve system capacity and system coverage. A node within the BS service range can be assigned a relay node depending on the network conditions and traffic requirements. Nodes close to each other are separated into different groups, and the BS serves the groups using the Round Robin scheduling policy to mitigate interference. Through Monte Carlo simulation techniques for both uplink and downlink, the authors show that the throughput of cell edge users can be improved from $150\%$ to $300\%$. The cell coverage can also be enlarged with significant data rates. 

Yang \textit{et al.} in \cite{yang_solving_2013} propose an architecture to setup D2D links for LTE-A based system, which is seldom considered by other researchers. This architecture includes a reference point between the D2D-enabled users to support proximity measurement, D2D channel state measurements, and D2D data transmission. A D2D bearer that \on{offloads} \off{takes off} traffic \off{load} from the Evolved Packet System (EPS) bearer is also included to provide the direct traffic path between users. The authors propose \on{ to include a function in the Packet data Gateway (P-GW) for proximity services.} \off{protocol entity} In addition, a protocol architecture is proposed to manage the D2D bearer and support D2D enhancement. Through an example, the authors present the detailed procedure to offload data from cellular-user links to D2D links.

Min \textit{et al.} in \cite{min_reliability_2011} try to improve the reliability of D2D communications through \emph{receive mode} selection. They consider three receive modes in cellular networks with underlay D2D communications: $(i)$ the D2D receiver decodes the desired signal directly while treating other signals as noise~\cite{annapureddy2009gaussian}; $(ii)$~the D2D receiver conceals other signals first and then decodes the desired signal~\cite{chung2007capacity}; and $(iii)$  the BS retransmits the interference from cellular communication to the D2D receiver. 
The last mode is proposed by the authors to improve the reliability of the D2D link. The paper investigates the outage probability under these three receive modes and provides close\on{d}-form results for computing outage probability. Each D2D user can separately calculate the outage probability for each receive mode using the closed-from formulas provided from the analysis. At each time instant, the users dynamically choose the best mode (i.e., the mode with the lowest outage probability). In order to reduce the energy consumption of the mobile device, the BS \off{can} perform\on{s} the outage probability calculations and send\on{s} the results to each user. However, this approach increases the computational overhead of the BS. 
Numerical results show that the outage probability can be improved by up to $99\%$ under the proposed receive method, which increases the reliability of D2D communications.

To ensure the reliability of cellular users, the authors of~\cite{kaufman2013ton} propose a scheme that does not cause outage for cellular users. They state that assuming D2D users have knowledge of location and channel state of cellular users is not feasible in a real system. Therefore, they design a distributed power control scheme that leverages a predefined interference margin of cellular users. Then, D2D users adjust their power level in such a way that their transmission does not exceed the interference margin of cellular users. D2D power adjustment can be done if the interference margin and estimating the channel gain between D2D user and the BS are known. The authors also propose to use distributed source routing algorithm to perform multi-hop D2D communication in the network. Simulation results indicate that the outage probability of D2D links reduces as the pathloss component \on{of the D2D link} increases.

Han \textit{et al.} in \cite{han_uplink_2012} consider the uplink channel reuse in a single-cell \off{cellular} network. The aim is to maximize the number of admitted D2D links while \off{minimize} \on{minimizing} the average interference caused by D2D links. The authors formulate the problem as a non-linear programming and design a heuristic algorithm based on the Hungarian algorithm \cite{Kuhn1956}. Their simulation results show that the performance of the proposed heuristic algorithm can be as good as the optimal solution. However, from the results we can see that the number of admitted D2D links under the proposed algorithm \on{does not increase} \off{cannot be increased} greatly as compared to a random D2D link allocation (\off{only at most} \on{maximum} one D2D link, which is less than $10\%$).

The authors of~\cite{pratas_low_2013} propose to use D2D communication to accommodate M2M communications in cellular networks. They state that M2M communications usually need low data rate but they are massive in numbers, \on{which leads to highly increased control overhead.} \on{Moreover, M2M communication is usually handled by a random medium access technique, which is susceptible to congestion and limited by number of contending users~\cite{Seo2011,Seo2012,Cheung2012,Tyagi2013,Joo2013}. } \off{Handling a large number of users creates overwhelming overhead on cellular networks.} Therefore, the authors \on{in \cite{pratas_low_2013}} propose to use \on{network-assisted} D2D communication among several machines and a cellular device. Next, the cellular device is used to relay M2M traffic to the BS. This approach can significantly reduce the overhead for the BS. The authors show via numerical simulations that their approach achieves $100\%$ gain over the scheme that does not make use of D2D communications.

The work in~\cite{du_compressed_2012} proposes a hybrid automatic repeat request (HARQ) for multicast in D2D enabled cellular networks. 
\off{The idea is to divide users into clusters, where in each cluster there is a cluster header (CH). The BS broadcasts frames to all devices within a cluster. Devices that fail to receive the frame send NACK} 
\off{to the CH via D2D links. The CH collects the feedback and sends compressed ACK/NACK to the BS. 
If some of the devices in the cluster have not received the frame correctly but at least one device has successfully received the frame, the CH will select a device to forward the frames to other devices.}
\on{The idea is to divide users into clusters, and have the BS broadcast packets to all devices within a cluster. In each cluster there is a Cluster Header (CH). A non-CH user that fails to receive the broadcast packet reports NACK to the CH via the D2D link. The CH can report the status of the broadcast transmission to the BS via a message stating one of the following states: $(i)$ an All\_ACK message that represents all the users within the cluster have successfully received the broadcast packet; $(ii)$ an All\_NACK message representing that all the users within the cluster have failed to receive the broadcast packet, so that the BS has to re-broadcast the packet; $(iii)$ a Self\_ACK message representing that the CH has successfully received the packet but at least another user has failed to receive it. The CH then transmits the packet to those that have failed to received it; and $(iv)$ a Self\_NACK message representing that the CH has failed to receive the frame but at least another user has successfully received it. Then the CH will choose a user that has received the packet to transmit the packet to those that have failed to receive it.}
This method highly reduces the frame loss ratio of the feedback (NACK/ACK) from devices compared to the method where each device sends ACK/NACK to the BS. Therefore, the performance of multicast is improved.

In~\cite{bao_dataspotting_2010}, D2D communications is used for content distribution in cellular networks. The authors propose a location-aware scheme which keeps track of the location of users and their requests. For example, if the BS receives a request from user $a$ for a content which is available in the cache of a nearby user $b$, it instructs user $b$ to send the content via D2D link to user $a$. \off{Therefore, the radio resources scheduled for the transmission of BS and user $a$ are saved because the BS does not have to re-transmit the same content. These resources can be scheduled to support the communication of BS and other users.}
\on{Using this method, the BS does not require to re-transmit a content which has been already transmitted. The amount of bandwidth saved with D2D communications can be used for future or pending transmissions. Using this approach, we can potentially reduce the transmission delay and increase the capacity of the network. It should be noted that keeping track of users' location and their cached traffic can lead to high control overhead. Moreover, the location tracking method should be optimized so that the battery of cellular device is not drained by the GPS.}

\begin{table*}[t!]
\centering
\caption{Summary of the literature proposing underlaying inband D2D}
\label{tb:underlay_inband}
\begin{tabular}{| l | l | l | l | l | l | l |}
\hline
Proposal & Analytical tools	& Platform & Direction & Use-case & Evaluation & \on{Achieved performance} \\
\hline \hline
\tabincell{l}{Improving spectrum\\efficiency \cite{peng_interference_2009,xu_effective_2010}\\
	\cite{janis_interference_2009, kaufman_cellular_2008, min_capacity_2011, 
	chen_downlink_2012}\\ 
 	\cite{yu_device_2012, doppler_device_2009, doppler_mode_2010, osseiran_advances_2009}\\	
 	\cite{zulhasnine_efficient_2010, pei_resource_2013, liu_optimal_2012, xu_resource_2012}}
	& \tabincell{l}{-Chen-Stein method\\ -Zipf distribution\\ -Integer/linear \\~programming\\
	-Mixed integer \\~nonlinear\\~programming\\ -Convex optimization\\ -Bipartite Matching\\ 
	-Kuhn-Munkres \\~algorithm\\ -Han-Kobayashi\\ -Newton's method\\ -Lagrangian\\~multipliers\\ 
	-Graph theory\\ -Auction algorithm\\ -Particle swarm \\~optimization}
	& \tabincell{l}{-WiMax\\ -CDMA\\ -LTE\\ -LTE-A} 
	& \tabincell{l}{-Uplink\\ -Downlink\\ -Uplink/\\~downlink}
	& \tabincell{l}{-Content \\~distribution\\ -File sharing\\ -Video/file\\~exchange}
	& \tabincell{l}{-Numerical \\simulation\\ -System-level \\simulation}
	& \tabincell{l}{\on{-System throughput can be improved} 
		\\\on{~from $16\%$ to $374\%$ compared with }
		\\\on{~conventional cellular networks under }
		\\\on{~common scenarios }
		\\\on{-Throughput can be improved up to}
		\\\on{~$650\%$ when D2D users are far away}
		\\\on{~from the BS}
		\\\on{-Number of admitted D2D users }
		\\\on{~can be increased up to $30\%$}} \\ \hline	
\tabincell{l}{Improving power\\efficiency \cite{xiao_qos_2011,jung_joint_2012}\\ 		
	\cite{belleschi_performance_2011,yu_power_2009}}
	& \tabincell{l}{-Heuristic algorithms\\ -Exhaustive search\\ -Linear programming}
	& \tabincell{l}{-LTE\\ -LTE-A\\ -OFDMA} & \tabincell{l}{-Uplink\\ -Downlink\\ 
	-Uplink/\\~downlink}
	&  & \tabincell{l}{-System-level \\simulation}
	& \tabincell{l}{\on{-Power efficiency can be improved} 
		\\\on{~from $20\%$ to $100\%$ compared with} 
		\\\on{~conventional cellular networks}} \\ \hline
\tabincell{l}{Improving performance\\ with QoS/power
	\\ constraints \cite{feng_device_2013,su_resource_2013}\\ 
	\cite{han_subchannel_2012,le_fair_2012, yu_power_2009, chia_resource_2011}}
	& \tabincell{l}{-Heuristic algorithms\\ -Bipartite Matching\\ -Kuhn-Munkres \\~algorithm}
	& \tabincell{l}{-LTE\\ -LTE-A} & \tabincell{l}{-Uplink\\ -Downlink}	
	& \tabincell{l}{-VOIP/FTP}
	& \tabincell{l}{-Numerical \\simulation \\-System-level \\simulation}  
	& \tabincell{l}{\on{-From $15\%$ to $70\%$ throughput gain} 
		\\\on{~with QoS constraint} 
		\\\on{- From $45\%$ to $500\%$ sum-rate gain} 
		\\\on{~with QoS/power constraint}} \\ \hline
\tabincell{l}{Improving fairness~\cite{xu_interference_2012}}\!\!\!
	& -Auction algorithm &  & -Downlink &  & 
	\tabincell{l}{-System-level \\simulation}
	& \tabincell{l}{\on{-A fairness index around $0.8\!\!$}} \\ \hline	
\tabincell{l}{Improving cellular\\ coverage \cite{vanganuru_system_2012}}
	&  & \tabincell{l}{-LTE\\-LTE-A} & \tabincell{l}{-Uplink/\\~downlink} &  & 
	\tabincell{l}{-Numerical \\simulation}
	& \tabincell{l}{\on{-Throughput of cell edge users can} 
		\\\on{~be improved up to $300\%$}
		\\\on{-Cell coverage is also enlarged} 
		\\\on{~up to $20\%$}} \\ \hline	
\tabincell{l}{Supporting setup of\\ D2D \cite{yang_solving_2013}}
	& -Protocol & -LTE-A & \tabincell{l}{-Uplink/\\~downlink} & 
	\tabincell{l}{-D2D link\\~setup} &
	& \tabincell{l}{}  \\ \hline
\tabincell{l}{Improving reliability\\ \cite{min_reliability_2011,kaufman2013ton}}
	& &	 &  &  & \tabincell{l}{-Numerical \\simulation}
	& \tabincell{l}{\on{-Outage probability reduces by~$99\%$}} \\ \hline
\tabincell{l}{Increasing the number\\ of concurrent D2D\\ links\cite{han_uplink_2012}}
	& \tabincell{l}{-Mixed-integer \\~nonlinear\\~programming \\ -Hungarian algorithm 
	\\ -Heuristic algorithm} & -LTE & -Uplink &  
	& \tabincell{l}{-System-level \\simulation}
	& \tabincell{l}{\on{-Number of admitted D2D links is} 
		\\\on{~increased up to $10\%$ compared to}  
		\\\on{~random D2D link allocation}} \\ \hline
\tabincell{l}{Offloading traffic \cite{bao_dataspotting_2010}}
	& &	 &  & \tabincell{l}{-Offloading\\~traffic} & \tabincell{l}{-System-level 
	\\simulation}
	& \tabincell{l}{} \\ \hline			
\tabincell{l}{Improving performance\\ of multicast~\cite{du_compressed_2012, 
	seppala_network_2011}} 
	& & \tabincell{l}{-LTE\\ -LTE-A} & \tabincell{l}{-Uplink/\\~downlink} & -Multicast
	& \tabincell{l}{-Numerical \\simulation\\ -System-level \\simulation}
	& \tabincell{l}{\on{-Frame loss ratio of feedback is} 
		\\\on{~reduced by $80\%$}} \\ \hline
\end{tabular}
\end{table*}

\on{\textbf{Summary:} In this subsection, the surveyed literature focused on various metrics and use-cases. Some employed advanced mathematical techniques such as  game theory to improve system performance in terms of spectrum utilization~\cite{cheng_resource_2012} or fairness~\cite{xu_interference_2012}. Using advanced mathematical techniques such as stochastic Lyapunov optimization~\cite{Neely2010} and dynamic programming might lead to increased complexity but they are indeed effective enhancement approaches which give the researchers insight to evaluation of other metrics such as queue stability and packet transfer time. The authors of~\cite{doppler_device_2009} provide the first protocol for signaling and other functionality in D2D-enabled networks. This helps greatly the researchers and engineers who plan to implement D2D in real world. Nevertheless, the evaluation scenario in~\cite{doppler_device_2009} can be enhanced to a more realistic setup. Other papers focus on exploiting new use-cases of D2D communication, such as multicast~\cite{du_compressed_2012} and content distribution~\cite{bao_dataspotting_2010}. Although most of these papers have not used advanced mathematical tools, their proposals lead to high performance gains. Moreover, D2D communication appears to be a viable candidate for applications such as proximity peer-to-peer gaming and social networking.
}

\on{Finally, a}  summary of the works on underlay D2D communication in cellular networks is provided in Table~\ref{tb:underlay_inband}, \on{in terms of metrics, use-cases, analytical tools, evaluation method, scope, and achieved performances.}

\section{Overlaying Inband D2D}
\label{s:overlay}
Different from the works reviewed in the previous subsection, the authors of~\cite{fodor_design_2012,li_device_2012,zhou_intracluster_2013} propose to allocate dedicated resources for D2D communications. This approach eliminates the concerns \off{of} \on{for} interference from D2D communications on cellular transmissions, but reduces the amount of achievable resources for cellular communications.
\begin{table*}[t!]
\centering
\caption{Summary of the literature proposing overlaying inband D2D}
\label{tb:overlay_inband}
\begin{tabular}{| l | l | l | l | l | l | l |}
\hline
Proposal & Analytical tools	& Platform & Direction & Use-case & Evaluation & 
\on{Achieved performance} \\
\hline \hline
\tabincell{l}{Increasing energy \\efficiency \cite{fodor_design_2012}}
	&  & -LTE & -Uplink &  &\tabincell{l}{-Numerical\\~simulation }
	& \tabincell{l}{\on{-Energy efficiency can be increased}
		\\\on{~from 0.8 bps/Hz/mW to 20 bps/Hz/mW}} \\ \hline	
\tabincell{l}{Improving spectrum \\efficiency \cite{li_device_2012}}
	& \tabincell{l}{-Convex\\~Optimization} & & -Uplink &  &\tabincell{l}{-Numerical\\~simulation }
	& \tabincell{l}{\on{-Cell throughput is improved by} 
		\\\on{~$40\%$ over underlay mode}} \\ \hline	
\tabincell{l}{Improving performance \\of multicast \cite{zhou_intracluster_2013}}
	&  &  & -Downlink & \tabincell{l}{-Video \\~transmission} & \tabincell{l}{-Numerical\\~simulation }
	&  \tabincell{l}{\on{-$90\%$ gain in bandwidth compared to}
		\\ \on{~the method using only one retransmitter}} \\ \hline
\end{tabular}
\end{table*}

In~\cite{fodor_design_2012}, Fodor {\it et al.} elaborate on challenges of D2D communications in cellular networks and suggest to control D2D communications from the cellular network. They claim that network assistance can solve the inefficiencies of D2D communications in terms of service and peer discovery, mode selection, channel quality estimation, and power control. In a conventional peer and service discovery method, D2D users should send beacons in short intervals and monitor multiple channels which is very energy consuming. However, this process can become more energy efficient if the BS regulates the beaconing channel and assists D2D users so that they do not have to follow the power consuming random sensing procedure. BS assistance also improves the scheduling and power control which reduces the D2D interference. The authors use simple Monte-Carlo simulation to evaluate the performance of D2D communications. The results show that D2D can increase the energy efficiency from $0.8$ bps/Hz/mW to $20$ bps/Hz/mW in the best case scenario \on{where the distance between D2D users is $10$m}.

The authors of~\cite{li_device_2012} propose incremental relay mode for D2D communication in cellular networks. In incremental relay scheme, D2D transmitters multicast to both the D2D receiver and BS. In case the D2D transmission fails, the BS retransmits the multicast message to the D2D receiver. The authors claim that incremental relay scheme improves the system throughput because the BS receives a copy of the D2D message which is retransmitted in case of failure. Therefore, this scheme reduces the outage probability of D2D transmissions. Although incremental relay mode consumes part of the downlink resources for retransmission, the numerical simulation results show that this scheme still improves the cell throughput by $40\%$ in comparison to underlay mode.

In~\cite{zhou_intracluster_2013}, D2D communication is used to improve the performance of multicast transmission in cellular networks. Due to wireless channel diversity, some of the \on{multicast group} members \off{of the multicast group} (i.e., cluster) may not receive the data correctly. The authors propose to use D2D communications inside the clusters to enhance the multicast performance. Specifically, after every multicast transmission, some of the members which manage to decode the message will retransmit it to those which could not decode the message. Unlike the prior work in~\cite{spinella2009integration}~and~\cite{fitzek2007cognitive} where there is only one predefined retransmitter, the number of retransmitters in~\cite{zhou_intracluster_2013} changes dynamically to maximize the spectral efficiency. The authors show via numerical simulations that their proposed algorithm consumes $90\%$  less spectrum resources in comparison to the scenario with only one retransmitter. 

\on{\textbf{Summary:} In this subsection, we surveyed the works which proposed to use dedicated resources for D2D communications. A BS-assisted scheduling and D2D power control was proposed in~\cite{fodor_design_2012} in order to reduce D2D interference. Differently, the authors of~\cite{li_device_2012} and~\cite{zhou_intracluster_2013} focus on relaying use-case of D2D. Specifically,~\cite{li_device_2012} proposes to use the BS as a relay (backup re-transmitter) for the D2D transmission and~\cite{zhou_intracluster_2013} uses multiple D2D users as relays (re-transmitters) for multicasting. Both methods proposed in~\cite{li_device_2012} and~\cite{zhou_intracluster_2013} have low complexity which makes them practical for real world scenarios. The algorithm proposed in~\cite{fodor_design_2012} is much more complex, and it exhibits very high performance when the maximal distance between D2D users is short. 
} 

A summary of the works on overlay D2D communication in cellular networks is provided in Table~\ref{tb:overlay_inband}.

\section{Outband D2D}
\label{s:outband}
In this section, we review the papers in which D2D communications occurs on a frequency band \on{that is} not overlapping \on{with} the cellular spectrum. Outband D2D is advantageous because there is no interference issue between D2D and cellular communications. Outband D2D communication can be managed by the cellular network (i.e., controlled) or it can operate on its own (i.e., autonomous).

\subsection{Controlled}
In works that fall under \off{the category of controlled outband D2D} \on{this category}, the authors propose to use the cellular network advanced management features to control D2D communication to improve the efficiency and reliability of D2D communications. They aim to improve system performance in terms of \emph{throughput, power efficiency, multicast}, and so on.

The authors of ~\cite{zhou_group_2013} propose to use ISM band for D2D communications in LTE. They state that simultaneous channel contention from both D2D and WLAN users can dramatically reduce the network performance. Therefore, they propose to group D2D users based on their QoS requirement and allow only one user per group to contend for the WiFi channel. The channel sensing between groups is also managed in a way that the groups do not sense the same channel at the same time. They show via simulation that their approach increases the D2D throughput up to $25\%$ in comparison to the scenario in which users contend for the channel individually.   

\begin{figure*} [t!]
\begin{center}
\vspace{4mm}
\includegraphics[scale=0.7]{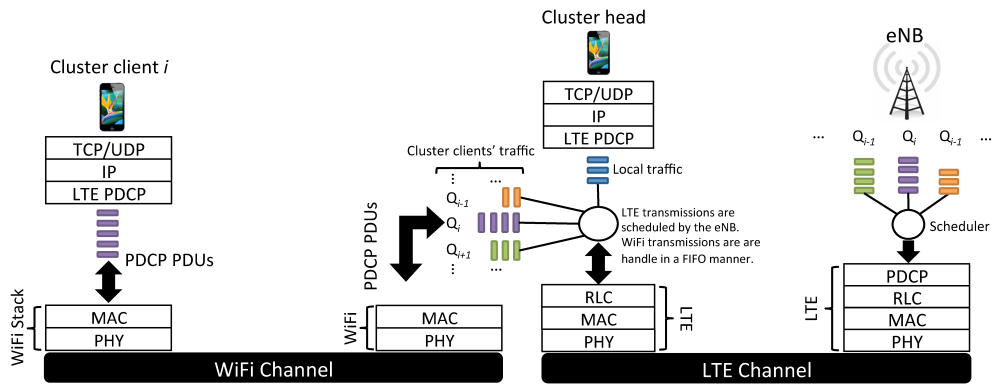}
\caption{Data flow between D2D users and the eNB (i.e., BS).} 
\label{fig:prot_stack}
\end{center}
\vspace{-5mm}
\end{figure*}

The authors of~\cite{arash2013MSWIM} and \cite{asadi2013energy} propose to use D2D communications for increasing the throughput and energy efficiency of cellular networks. The authors propose to form clusters among cellular users who are in range for WiFi communication. After the cluster is formed, only the cluster member with the highest cellular channel quality (i.e., cluster head) communicates with the BS. The cluster head is also responsible to forward the cellular traffic of its clients (i.e., other users who belong to the same cluster) to the BS. The authors provide an analytical model to compute the throughput and power consumption for the proposed scheme. The advantages of this scheme are threefold: $(i)$ the spectral efficiency increases because the cluster head has the highest channel quality among the cluster members which corresponds to transmissions with high Modulation and Coding Schemes (MCS); $(ii)$ the energy efficiency is increased because the cluster clients can go to cellular power saving mode; $(iii)$ the fairness can be increased because the cellular resources is distributed among cluster members in a way that users with poor channel quality are not starved.  The authors show via numerical simulation that D2D communications improve throughput and energy efficiency with respect to classical Round Robin schedulers by $50\%$ and $30\%$, respectively. The results also show that the proposed scheme can achieve almost perfect fairness. 

Furthermore, the work in~\cite{arash2013wireless} provides a protocol for \off{a} \on{the} D2D communication scheme proposed in~\cite{arash2013MSWIM}. The authors first elaborate on required modification \off{to} \on{for} messaging and signaling procedures of LTE and WiFi Direct technologies. Next, they define a protocol stack to connect the two technologies, as illustrated in Fig.~\ref{fig:prot_stack}.  \on{The protocol stack will be further elaborated in Section~\ref{s:protocols}.} This paper sheds light on different aspects of integrating LTE and WiFi Direct such as channel quality feedback, scheduling, security, \off{and} etc. Via a home grown LTE simulator, the authors show that the proposed scheme in~\cite{arash2013MSWIM} can improve the packet delay of a Round Robin scheduler by up to $50\%$ and it can guarantee delays less than $10$ ms with $99\%$ probability.

Golrezaei {\it et al.}~\cite{golrezaei_base_2012,golrezaei2012device} 
point out the similarities among video content requests of cellular users. They propose to cache the popular video files (i.e., viral videos) on smartphones and exploit D2D communications for viral video transmissions in cellular networks. They partition each cell into clusters (smaller cells) and cache the non-overlapping contents within the same cluster. When a user sends a request to the BS for a certain content, the BS checks the availability of the file in the cluster. If the content is not cached in the cluster, the user receive\off{d}\on{s} the content directly from the BS. If the content is locally available, the user receives the file from its neighbor in the cluster over unlicensed band (e.g., via WiFi). The authors claim that their proposal improves the video throughput by one or two orders of magnitude.

The authors of~\cite{ji_wireless_2013} propose a method to improve video transmission in cellular networks using D2D communications. This method exploits the property of asynchronous content reuse by combining D2D communication and video caching on mobile devices. Their objective is to maximize per-user throughput constrained to the outage probability (i.e., the probability that a user's demand is unserved). They assume devices communicate with each other with fixed data rate and there is no power control over the D2D link. Through simulations, the authors show that their proposed method outperforms the schemes with conventional unicast video transmission as well as the coded broadcasting~\cite{maddah2012fundamental}. The results show that their proposed method can achieve at least \off{$100$ times and $10$ times}  \on{$10000\%$ and $1000\%$} throughput gain over the conventional and coded broadcasting methods, respectively, when the outage probability is less than $0.1$.

Cai \textit{et al.} \cite{Cai2013} propose a scheduling algorithm to exploit both time-varying channel and users' random mobility in cellular networks. They consider a scenario where the BS broadcasts deadline-based content to different group of users. Users move randomly within the cell and users of the same group are assumed to be able to communicate directly at high rate when they are close to each other. Therefore, users can exchange all the content within their current lists during a contact period. During each slot, the BS dynamically selects a group of users to broadcast content to at a chosen service rate, based on the scheduling algorithm employed. If the service rate is too high for some users to successfully receive the content, these users will exploit the D2D communication to fetch content from nearby users in the near future. The authors formulate the scheduling problem with objective to maximize the group utility function. Next, they solve the maximization problem under the assumption of statistically homogeneous user mobility, and then extend it to the heterogeneous scenarios. Simulation results show that the proposed scheduling algorithm can improve the system throughput \off{by} \on{from} $50\%$ to $150\%$, compared to the scheduling algorithm without D2D communications.

\on{\textbf{Summary:} The works addressed in this subsection focus on various use-cases of D2D communication. The authors in~\cite{arash2013MSWIM} and~\cite{asadi2013energy} use clustering and game theory to boost the throughput performance as well as energy efficiency and fairness. For the first time, they designed a detailed protocol for outband D2D communications in~\cite{arash2013wireless}. 
The work in~\cite{golrezaei_base_2012,golrezaei2012device,ji_wireless_2013} and~\cite{Cai2013} aim to improve the performance of content distribution. The methods proposed in~\cite{golrezaei_base_2012, golrezaei2012device} are simple, while that of~\cite{ji_wireless_2013} is more complex. The performance of both methods is evaluated to be good. In addition to content distribution, the authors in~\cite{Cai2013} also consider user mobility and deadline-based content, for which they provide comprehensive evaluations under a realistic simulation setup including real-time video transmission.
}

\subsection{Autonomous}
\begin{figure}[!t]
  \centering
  \includegraphics[width=88mm]{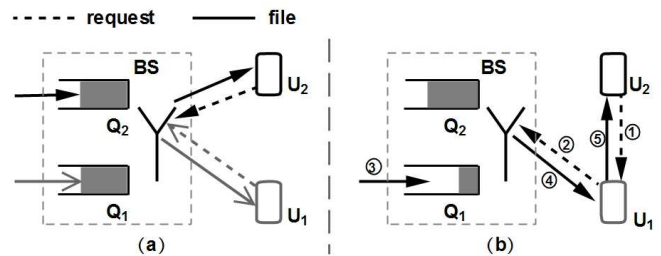}
  \caption{An example of the BS-transparent traffic spreading: (a) No traffic spreading; (b) Traffic spreading from $U_2$ to $U_1$.}
  \label{fig:wang2013wowmom_scheme}
\end{figure}

Autonomous D2D communication is usually motivated by reducing the overhead of cellular networks. It does not require any changes at the BS and can be deployed easily. Currently, there are very few works on this category. Wang \textit{et al.}~\cite{wang2013WOWMOM} propose a downlink BS-transparent dispatching policy where users spread traffic requests among each other to balance their backlogs at the BS, as shown in Fig.~\ref{fig:wang2013wowmom_scheme}. They assume that users' traffic is dynamic, i.e., the BS does not always have traffic to send to all the users at any time. \on{They illustrate the dispatching policy by considering a scenario with two users, $U_1$ and $U_2$ being served by the BS. The queues $Q_1$ and $Q_2$ depict the numbers of files at user's BS queues. In Fig.~\ref{fig:wang2013wowmom_scheme}(a), since the queues at the BS are balanced, the dispatchers at each user would detect that traffic spreading is not beneficial. Thus users send their new requests to the BS directly. In Fig.~\ref{fig:wang2013wowmom_scheme}(b), there are more files in $Q_2$ than $Q_1$. The dispatcher of $U_2$ would detect that traffic spreading is beneficial, because in the near future $Q_1$ would be empty and thus the opportunistic scheduling gain is lost. Therefore, $U_2$ asks $U_1$ to forward its new file requests to the BS. After receiving the corresponding files from the BS, $U_1$ forwards them to $U_2$.} 
This dispatching policy is user-initiated (i.e., it does not require any changes at the BS) and works on a per-file basis. \off{It} \on{This policy} exploits both the time-varying wireless channel and users' queueing dynamics at the BS \off{and the aim is} \on{in order} to reduce average file transfer delays seen by the users.  The users perceive their channel conditions to the BS (i.e., cellular channel conditions) and share them among each other. The authors formulate the problem of determining the optimal file dispatching policy under a specified tradeoff between delay performance and energy consumption as a Markov decision problem. Next, they study the properties of the corresponding optimal policy in a two-user scenario. A heuristic algorithm is proposed which reduces the complexity in large systems by aggregating the users. 
\off{Based on realistic Rayleigh fading channels, the authors provide simulation results demonstrating that file transfer delays can be reduced by up to 50\% using the proposed methodology and that significant gains (up to 78\% of the gain) are typically achieved at only $20\%$ of the power expenditure of the performance-centric algorithm, which achieves the best performance at high power expenditure.}
\on{The simulation results demonstrate that the file transfer delays can be reduced by up to $50\%$ using the proposed methodology. In addition, their proposal consumes $80\%$ less power than performance-centric algorithms while achieving significant gains (up to $78\%$).}
\begin{table*}[t!]
\centering
\caption{Summary of the literature proposing outband D2D}
\label{tb:outband}
\begin{tabular}{| l | l | l | l | l | l | l |}
\hline
Proposal & Analytical tools	& Platform & Direction & Use-case & Evaluation & \on{Achieved performance}\\
\hline \hline
\tabincell{l}{Improving throughput\\ of video distribution\\ \cite{golrezaei_base_2012, 
	golrezaei2012device}} & \tabincell{l}{-Game theory\\ -Chen-Stein method}  & & -Downlink 
	& -Content distribution & \tabincell{l}{-Numerical \\~simulation} 
	& \tabincell{l}{\on{-Video throughput is }
		\\\on{~improved by up to two }
		\\\on{~orders of magnitude}} \\ \hline	
\tabincell{l}{Reducing channel\\ sensing overhead~\cite{zhou_group_2013}}
	&  & -LTE & & -Relaying & \tabincell{l}{-System-level\\~simulation}
	& \tabincell{l}{\on{-Throughput is improved} 
		\\\on{~by up to $25\%$}} \\ \hline	
\tabincell{l}{Improving  throughput,\\ energy efficiency, and \\fairness~\cite{arash2013MSWIM, 
	asadi2013energy, Cai2013}} & -Game theory  & \tabincell{l}{-LTE\\ -CDMA} & -Downlink 
	& \tabincell{l}{-Relaying\\ -Video transmission} 
	& \tabincell{l}{-Numerical\\ ~simulation} 
	& \tabincell{l}{\on{-Throughput and energy} 
		\\\on{~efficiency are improved} 
		\\\on{~by $50\%$ and $30\%$ over} 
		\\\on{~classical Round Robin} 
		\\\on{~scheduler, respectively}} \\ \hline
\tabincell{l}{Designing a protocol\\ for outband D2D \\communications~\cite{arash2013wireless}} 
	&   & -LTE & \tabincell{l}{ -Downlink \\-Uplink}
	&-Relaying  & \tabincell{l}{-System-level \\~simulation}
	& \tabincell{l}{\on{-$50\%$ delay improvement}
		\\\on{~compared to Round Robin}
		\\\on{~scheduler }} \\ \hline	
\tabincell{l}{Improving video\\ transmission~\cite{ji_wireless_2013}}
	&  & -LTE & -Downlink & -Video transmission 
	& \tabincell{l}{-System-level \\~simulation} 
	& \tabincell{l}{\on{-Throughput is improved by}
		\\\on{~$10000\%$ and $1000\%$ over }
		\\\on{~conventional and coded}
		\\\on{~broadcasting methods, }
		\\\on{~respectively}} \\ \hline
\tabincell{l}{Reducing average file\\ transfer delay~\cite{wang2013WOWMOM}}
	& \tabincell{l}{-Dynamic programming\\ -Heuristic algorithm\\ 
	-Distributed algorithm\\ -Queueing theory} & -LTE & -Downlink
	& \tabincell{l}{-Web browsing\\ -HTTP live streaming} 
	& \tabincell{l}{-System-level \\~simulation}
	& \tabincell{l}{\on{-Average file transfer delay }
		\\\on{~is reduced by up to $50\%$ }
		\\\on{~compared to methods}
		\\\on{~without traffic spreading}} \\ \hline
\end{tabular}
\end{table*}

A summary of the works on outband D2D communication in cellular networks is provided in Table~\ref{tb:outband}.

%

\on{
\section{Proposed D2D Protocols}
\label{s:protocols}

The majority of researchers have addressed issues such as interference, resource allocation, power allocation, and  so on. Only a few researchers propose protocols for D2D communications. In particular, the authors of~\cite{raghothaman2013ICNC}~and~\cite{arash2013wireless} propose a protocol stack for inband and outband D2D communication, respectively. 

In~\cite{raghothaman2013ICNC}, the authors describe the required architectural and protocol modification in the current cellular standards to adapt inband D2D communication. The main architectural modification consists in adding a D2D server inside or outside the core network. In case the D2D server is placed outside the core network, it should have interfaces with Mobility Management Entity (MME), Policy and Charging Rules Function (PCRF), peer D2D servers, and application servers. The D2D server is expected to handle functionalities such as device identifier allocation, call establishment, UE capability tracking, service support, and mobility tracking. Fig.~\ref{fig:inband_architecture} illustrates the architecture which was described above. 
The authors also propose a protocol stack in which D2D pairs have extra PHY, MAC, Radio Link Control (RLC), and Packet Data Convergence Protocol (PDCP) layer for direct communication. This means that UEs retain their cellular connectivity while communicating over D2D link. 

\begin{figure}[!t]
  \centering
  \includegraphics[width=85mm]{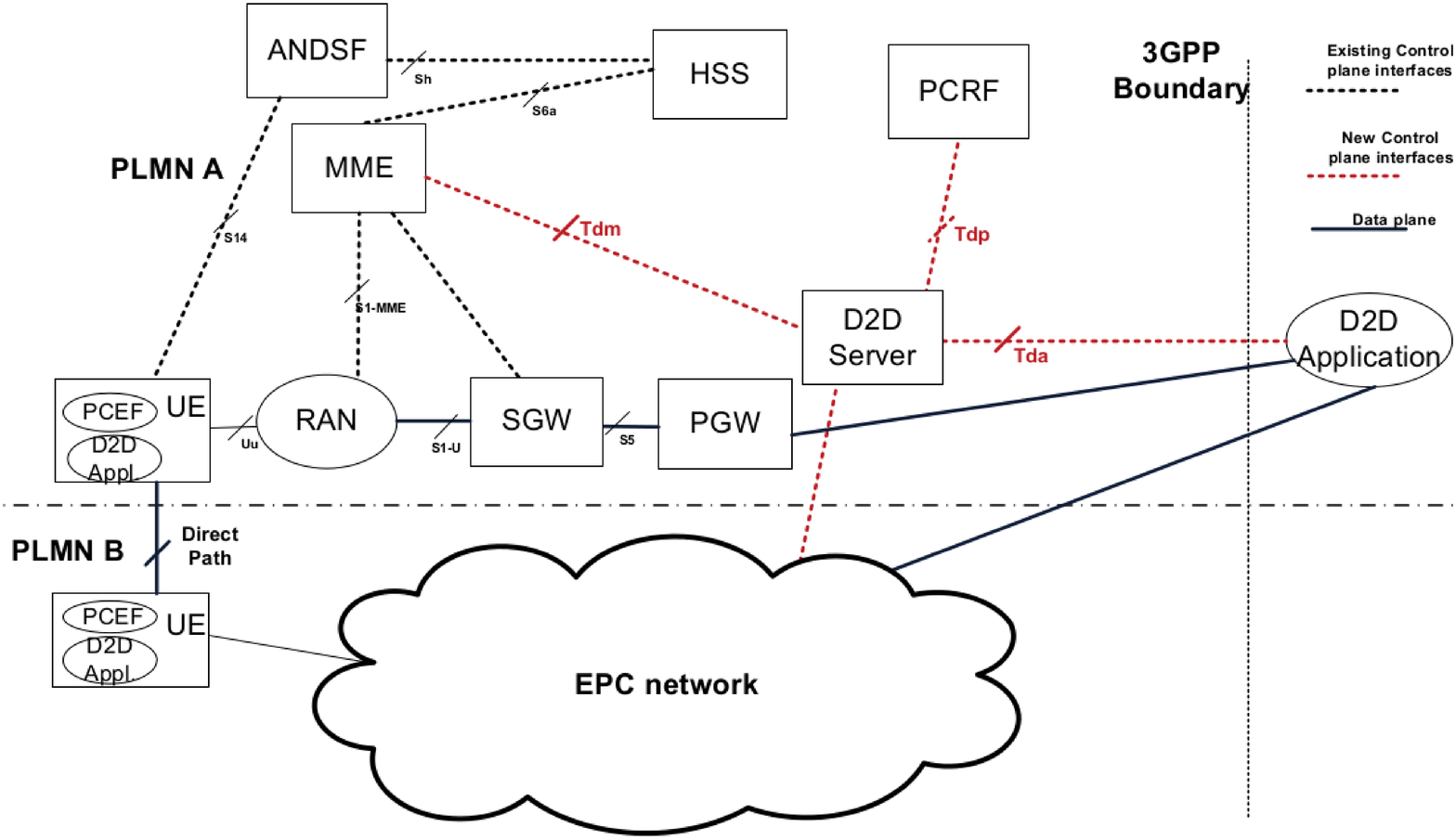}
  \caption{The illustration of proposed D2D architecture in~\cite{raghothaman2013ICNC}.}
  \label{fig:inband_architecture}
\end{figure}

\begin{figure*}[!t]
  \centering
  \includegraphics[scale=0.6]{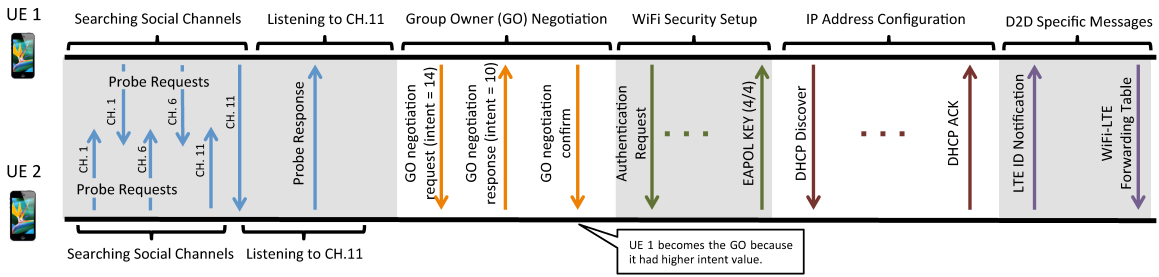}
  \caption{D2D device discovery and connection establishment procedure.}
  \label{fig:wifi-direct}
\end{figure*}

The authors of~\cite{arash2013wireless} elaborate on feasibility of outband D2D communications in LTE-based systems. As mentioned before, the target of their proposal is opportunistic packet relaying. To this aim, the authors provide a protocol stack which connects LTE and WiFi Direct protocols (see Fig.~\ref{fig:prot_stack}). The authors propose to encapsulate LTE PDCP Packet Data Units (PDUs) into WiFi packet(s) and transmit it (them) over WiFi to the D2D receiver. If the D2D receiver needs to relay the packet to the eNB, it simply extracts the LTE PDCP PDU from WiFi and processes it through RLC, MAC and PHY layers, as shown in~Fig.~\ref{fig:prot_stack}. In addition to providing a protocol stack, important procedures such as device discovery, D2D registration, connection establishment, default/dedicated bear setup, mobility management, CSI reporting, scheduling, security are also addressed in this paper. While addressing all these issues, the authors try to minimize the modification to the existing protocols. For example, the discovery phase and connection establishment is very similar to WiFi Direct standard defined procedure. The main difference is the addition of an extra phase (i.e., D2D Specific Messages, see Fig.~\ref{fig:wifi-direct} ) in order to exchange LTE IDs between D2D users. For detailed description of the protocol refer to~\cite{arash2013wireless}. None of the proposed protocols for D2D communication comment on when/how to activate D2D mode. Thus, this remains an open research problem to be solved in future.

As mentioned earlier, two 3GPP Working Groups are also investigating the ProSe use-cases~\cite{3GPPTR22.803} in LTE and required protocol/architecture enhancements~\cite{3GPPTR23.703} to accommodate such use-cases.
ProSe communication supports two types of data path: direct mode and locally-routed. In direct mode, two UEs exchange data directly with each other. In locally-routed, data between UEs is routed locally via the eNB(s). Both of these are different from the data path specified in current LTE standard, where the Serving Gateway/Packet data network Gateway (SG/PGW) is involved. Besides, control path in ProSe communication has more choices. If two UEs using the ProSe communication are served by the same eNB, the system can decide to perform control information between UE, eNB and EPC. The UEs can also exchange control signaling directly with each other to minimize signaling modification. If two UEs involved in the ProSe communication are served by different eNBs, the system can decide to perform control information between UE, eNB and EPC. In additional, the eNBs can coordinate with each other directly for radio resource management, and the UEs can communicate directly to exchange control signaling. 
The 3GPP also defines other aspects of ProSe communication, such as ProSe direct discovery, roaming, support for public safety service and support for WLAN direct communication, etc. Details of these aspects can be found in~\cite{3GPPTR23.703}. Based on the above mentioned schemes, the 3GPP proposes tens of use-cases~\cite{3GPPTR22.803}, such as ProSe-enabled UEs discover other ProSe-enabled UEs (which can be used for social networking), supports large number of UEs in dense environment (which can be used for city parking service), establishes ProSe-assisted WLAN direct communications (which can be used for cellular traffic offloading), and so on.

}

\section{Discussions and Future Work}
\label{s:discussions}

So far we have reviewed the available literature on D2D communications in cellular networks. In this section, we will shed light on some important factors such as common assumptions, scope of the works, and common techniques. 

\subsection{Common Assumptions} 

\on {Most of the papers in literature assume the BS is aware of the instantaneous CSI of cellular and/or D2D links, e.g.,~\cite{min_capacity_2011,xiao_qos_2011,feng_device_2013, yu_device_2012,yu_power_2009,su_resource_2013,han_uplink_2012}. This assumption is essential because their proposed solutions need the BS's participation to make scheduling decision for cellular and D2D users. Alternatively, when the D2D users decide on the their transmission slots, the common assumption is that D2D users are aware of the cellular and D2D links. On the other hand, there are also papers such as~\cite{yu_performance_2009} and \cite{wang2013WOWMOM} that assume the BS or D2D users are only aware of the statistical CSI of the links. With this assumption, the large overhead for reporting instantaneous CSI can be avoided.
To mitigate possible interference from D2D transmission to cellular transmission, \cite{kaufman_cellular_2008} assumes that D2D users are aware of minimum interference threshold of cellular users. With the latter assumptions, the D2D users can opportunistically choose the transmission slots in which they do not interfere with the cellular users. 

The proposals which involve in clustering users commonly assume that the cluster are far enough so that there is no or negligible interference among different clusters, e.g.,~\cite{du_compressed_2012, cheng_resource_2012,asadi2013energy,arash2013MSWIM}. This assumption may not hold in populated areas or dense deployments. A very interesting observation from the reviewed literature is that the majority of papers assume that the BS or D2D users always have traffic to send, therefore they use throughput as a common metric. However, the authors of~\cite{wang2013WOWMOM} consider a scenario with dynamic traffic load and evaluate the average file transfer delay under their proposed traffic spreading mechanism. Since the latter assumption is more realistic, it would be interesting to see the performance of the aforementioned works under dynamic traffic flows. }  

\subsection{Inband or Outband?}
Majority of the papers propose to reuse the cellular resources for D2D communications (i.e., inband)~\cite{kaufman_cellular_2008,doppler_device_2009,doppler_device_2009A,doppler_mode_2010,doppler_advances_2011}. However, outband communication is attracting more and more attention in the past few years~\cite{arash2013MSWIM,asadi2013energy,wang2013WOWMOM}.  Before comparing the two approaches, we summarize the advantages and disadvantages of each approach. 

\begin{table*}[t!]
\centering
\caption{Advantages and disadvantages of  different types of D2D communications}
\label{tb:adv-disadv}
\begin{tabular}{|l|c|c|c|c|}
\cline{2-5}
\multicolumn{1}{c|}{}							&\multicolumn{2}{|c|}{Inband} 			&\multicolumn{2}{|c|}{Outband}	 		\\
\cline{2-5}
\multicolumn{1}{c|}{}							&Underlay		&Overlay			&Controlled		&Autonomous		\\	
\hline
\hline
Interference between D2D and cellular users 		&$\checkmark$		&$\times$			&$\times$			&$\times$			\\
\hline
Requires dedicated resources for D2D users 		&$\times$			&$\checkmark$		&$\times$			&$\times$			\\
\hline
Controlled interference environment 				&$\checkmark$		&$\checkmark$		&$\times$			&$\times$			\\
\hline
Simultaneous D2D and cellular transmission		&$\times$			&$\times$			&$\checkmark$		&$\checkmark$		\\
\hline
Requires inter-platform coordination				&$\times$			&$\times$			&$\checkmark$		&$\checkmark$		\\
\hline
Requires devices with more than one radio interface&$\times$			&$\times$			&$\checkmark$		&$\checkmark$		\\
\hline
Introduces extra complexity to scheduler			&$\checkmark$		&$\checkmark$		&$\checkmark$		&$\times$			\\
\hline
\end{tabular}
\end{table*}		

{\bf Inband.} Inband D2D is advantageous in the sense that: $(i)$ underlay D2D increases the spectral efficiency of cellular spectrum by exploiting the spatial diversity; $(ii)$ any cellular device \off{are} \on{is} capable of using inband D2D communication (the cellular interface usually does not support outband frequencies); and $(iii)$ QoS management is easy because the cellular spectrum can be fully controlled by the BS. The disadvantages of inband D2D communications are: $(i)$ cellular resources might be wasted in overlay D2D; $(ii)$ the interference management among D2D and cellular transmission in underlay is very challenging; $(iii)$ power control and interference management solutions usually resort to high complexity resource allocation methods; and $(iv)$ a user cannot have simultaneous cellular and D2D transmissions. 
It appears that underlay D2D communication is more popular than overlay. The authors who propose to use overlay D2D usually try to avoid the interference issue of underlay~\cite{fodor_design_2012,li_device_2012,zhou_intracluster_2013}. However, allocating dedicated spectrum resources to D2D users is not as efficient as underlay in terms of spectral efficiency. We believe that the popularity of underlay D2D is due to its higher spectral efficiency. 

{\bf Outband.} This type of D2D communications has merits such as: $(i)$ there is no interference between cellular and D2D users; $(ii)$ there is no need for dedicating cellular resources to D2D spectrum like overlay inband D2D; $(iii)$ the resource allocation becomes easier because the scheduler does not require to take the frequency, time, and location of the users into account; \on{and} $(iv)$ simultaneous D2D and cellular communication is feasible. Nevertheless outband D2D \off{had} \on{has} some disadvantages which are: $(i)$ the interference in unlicensed spectrum is not in the control of the BS; $(ii)$ only cellular devices with two radio interfaces (e.g., LTE and WiFi) can use outband D2D communications; $(iii)$ the efficient power management between two wireless interfaces is crucial otherwise the power consumption of the device can increase; \on{and} $(iv)$ packets (at least the headers) need to be decoded and encoded because the protocols employed by different radio interfaces are not the same. 

Although the literature on inband D2D is wider than that of outband, it seems that researchers have started to explore the advantages of outband D2D and they are considering it as a viable alternative to inband D2D. We believe that with the evolutionary integration of smartphones in phone market, \on{the} majority of \off{the} mobile devices \off{are} \on{will be} equipped with more than one wireless interface which makes it possible to implement outband D2D schemes. Moreover, the standards such as 802.21~\cite{802.21} are looking into handover to and from different platforms (e.g., WiMAX and LTE ) which could significantly reduce the complexity of coordination between different wireless interfaces in outband D2D.  Table~\ref{tb:adv-disadv} summarizes the above mentioned merits and disadvantages.

\subsection{Maturity of D2D in Cellular Networks}
We believe D2D communication in cellular networks is a relatively young topic and there is a lot to be done/explored in this field.
We support this belief by looking into the analytical techniques and evaluation methods which are used in the available literature. 

{\bf Analytical techniques.}
In comparison to other fields such as opportunistic scheduling~\cite{asadi2013survey}, the number of techniques used in the literature and their popularity is very low. The majority of the literature only proposes ideas, architectures, or simple heuristic algorithms. 
Some of the papers formulate their objectives as optimization problems but leave them unsolved due to NP-hardness.
Therefore, we believe there is room for investigating optimal solutions for interference coordination, power management, and mode selection.  Table~\ref{tb:math} summarizes the mathematical techniques used in the D2D related literature. 

\begin{table}[h!]
\centering
\caption{Analytical tools used in the literature}
\label{tb:math}
\begin{tabular}{|l|c|}
\hline
Tools 							& { Related literature}\\
\hline
Discrete Time Markov chain			& \cite{akkarajitsakul_mode_2012} \\
Merge and split algorithm			& \cite{akkarajitsakul_mode_2012} \\
Distributed algorithms				& \cite{wang2013WOWMOM,akkarajitsakul_mode_2012, 
	han_uplink_2012}\\
Coalitional game theory				& \cite{akkarajitsakul_mode_2012}\\
Poisson point process				& \cite{lin_comprehensive_2013} \\
Queueing theory						& \cite{wang2013WOWMOM} \\
Alzer's inequality					& \cite{lin_comprehensive_2013} \\
Fubini's theorem					& \cite{lin_comprehensive_2013} \\
Laplace transform					& \cite{lin_comprehensive_2013,Erturk_distributions_2013} \\
Slivnyak's theorem					& \cite{lin_comprehensive_2013} \\
Heuristic algorithm					& \cite{wang2013WOWMOM,xiao_qos_2011, 
	han_uplink_2012} \\
Convex Optimization					& \cite{li_device_2012,liu_optimal_2012} \\
Chen-Stein Method					& \cite{golrezaei2012device} \\
Maximum Weight Bipartite Matching	& \cite{feng_device_2013}\\
Kuhn-Munkres algorithm				& \cite{feng_device_2013}\\
Han-Kobayashi						& \cite{yu_device_2012} \\
Jensen's Inequality					& \cite{Erturk_distributions_2013} \\
Mixed integer nonlinear programming	& \cite{zulhasnine_efficient_2010}\\
Integer programming					& \cite{le_fair_2012}\\
Linear programming					& \cite{belleschi_performance_2011, le_fair_2012}\\
Non-linear programming				& \cite{han_uplink_2012} \\
Dynamic programming					& \cite{wang2013WOWMOM} \\
Newton's method						& \cite{le_fair_2012} \\
Lagrangian multipliers				& \cite{le_fair_2012} \\
Graph theory						& \cite{zhang_interference_2013} \\
Auction algorithms					& \cite{xu_interference_2012,xu_resource_2012} \\	
Exhaustive search 					& \cite{jung_joint_2012} \\
Geometrical probability				& \cite{kaufman2013ton}\\
Evolution theory					& \cite{cheng_resource_2012} \\
Particle swarm optimization			& \cite{su_resource_2013} \\
Sub-gradient algorithm				& \cite{han_subchannel_2012} \\
Hungarian algorithm					& \cite{han_uplink_2012} \\
\hline
\end{tabular}
\end{table}

{\bf Evaluation method.}
Another metric for maturity of a field is the evaluation method. The more realistic the evaluation method, the more mature the study of that field. Table~\ref{tb:eval} shows different evaluation methods used in the literature. As we can see, majority of the papers use numerical evaluation and some use simple home-grown simulators. There is no paper using experimental evaluation. This is mainly due to the fact that experimental testbeds for cellular network are extremely costly and do not have support for D2D yet. The literature rarely uses popular network simulators such as NS3~\cite{ns3}, OPNET~\cite{opnet}, Omnet++~\cite{varga2007omnet++}.  In turn, currently available network simulators do not support D2D communications. 

\begin{table}[h!]
\centering
\caption{Evaluation methods in the literature}
\label{tb:eval}
\begin{tabular}{|l|c|}
\hline
Evaluation method		& {Related literature}\\
\hline
Numerical simulation	& \cite{min_capacity_2011, lin_comprehensive_2013, du_compressed_2012, 
	golrezaei_base_2012, kaufman_cellular_2008, doppler_device_2009, feng_device_2013, 
	yu_device_2012} \\
	& \cite{Erturk_distributions_2013, fodor_design_2012, chen_downlink_2012, le_fair_2012, 
	janis_interference_2009, zhou_intracluster_2013, pratas_low_2013, doppler_mode_2010} \\
	& \cite{cheng_resource_2012,pei_resource_2013,chia_resource_2011,han_subchannel_2012, 
	golrezaei2012device, zhang_interference_2013, yu_performance_2009, yu_power_2009} \\
	& \cite{yu_performance_2009,min_reliability_2011,xu_resource_2012,su_resource_2013, 
	vanganuru_system_2012,xu_resource_2012,su_resource_2013, vanganuru_system_2012} \\ \hline
System-level simulation & \cite{zulhasnine_efficient_2010, belleschi_performance_2011, 
	zhou_group_2013, peng_interference_2009,xu_resource_2012,liu_optimal_2012, 	
	doppler_device_2009A,janis_interference_2009} \\
	& \cite{xiao_qos_2011,rasmussen2000matrix,akkarajitsakul_mode_2012, ji_wireless_2013, 
	jung_joint_2012,bao_dataspotting_2010, seppala_network_2011, koskela_clustering_2010} \\
	& \cite{zhu_qos_2011,wang2013WOWMOM,han_uplink_2012} \\
\hline
Experiment	& No experimental study\\			
\hline
\end{tabular}
\end{table}


\subsection{How Far Is D2D from a Real World Implementation?}
Although D2D communication is not mature yet, it is already being studied in the 3GPP standardization body~\cite{3GPPTR22.803,3GPPTR23.703}. \on{ 3GPP recently decided that the focus of D2D in LTE would be on public safety networks~\cite{Lin2013ComMag}. } Moreover, Qualcomm has shown interests in this technology and they also built a prototype for D2D communications in cellular network which can be used in different scenarios such as social networking, content sharing, and so on~\cite{corson_toward_2010}. This confirms that D2D communication is not only a new research topic in academia, but also that there is interest for such a technology in the industry. \on{There are various obstacles to implement D2D in cellular network. For example, the operators are used to having control of their spectrum and the way it is used. As a result, a successful D2D implementation should allow D2D communications in a manner that operators are not stripped off their power to control their network. Moreover, there are physical challenges such as suitable modulation format and CSI acquisition which should be addressed efficiently. } Therefore, we believe that D2D communications will become an essential part of cellular communications in the next few years. 

\on{
\subsection{D2D Implementation Challenges in Real World}
Although D2D communication triggered a lot of attention and interest in academia, industry, and standardization bodies, it is not going to be integrated into the current communication infrastructure until the implementation challenges are resolved. Here, we explain some of the major challenges faced by D2D communications. 

{\bf Interference management.}  Under inband D2D communication, UEs can reuse uplink/downlink resources in the same cell. Therefore, it is important to design the D2D mechanism in a manner that D2D users do not disrupt the cellular services.  Interference management is usually addressed by power and resource allocation schemes, although the characteristics of D2D interference are not well understood yet.

{\bf Power allocation.} In inband D2D, the transmission power should be properly regulated so that the D2D transmitter does not interfere with the cellular UE communication while maintaining minimum SINR requirement of the D2D receiver. In outband D2D, the interference between D2D and cellular user is not of concern. Therefore, power allocation may seem irrelevant in outband D2D. However, with increased occupancy of ISM bands, efficient power allocation becomes crucial for avoiding congestion,  collision issues, and inter-system interference.

{\bf Resource allocation.} This is another important aspect of D2D communication specially for inband D2D. Interference can be efficiently managed if the D2D users communicate over resource blocks that are not used by nearby interfering cellular UEs. Resource allocation for outband D2D simply consists in avoiding ISM bands which are currently used by other D2D users, WiFi hotspots, etc.  

{\bf Modulation format.} This is one of the challenges which is rarely addressed by researchers. The existing LTE UEs use an OFDMA receiver in downlink and a SC-FDMA for uplink transmission. Thus, for using downlink (resp. uplink) resources, the D2D UE should be equipped with OFDMA transmitter (resp. SC-FDMA receiver)~\cite{Lin2013ComMag}. 

{\bf Channel measurement.} Accurate channel information is indispensable to perform efficient interference management, power allocation, and resource allocation. Conventional cellular systems only need the downlink channel information from UEs and the uplink channel information is readily computed at the base station. Unfortunately, D2D communication requires information on the channel gain between D2D pairs, the channel gain between D2D transmitter and cellular UE, and the channel gain between cellular transmitter and D2D receiver. The exchange of such extra channel information can become an intolerable overhead to the system if the system needs instantaneous CSI feedback. The trade-off between accuracy of CSI and its resulting overhead is to be further investigated. 

{\bf Energy consumption.} D2D communication can potentially improve the energy efficiency of the UE. However, this highly depends on the protocol designed for device discovery and D2D communication. For example, if the protocol forces the UE to wake up very often to listen for pairing requests or to transmit the discovery messages frequently, the battery life of the UE may significantly reduces. The trade-off between UE's power consumption and discovery speed of the UEs should be better studied.

{\bf HARQ.} Considering the complexity of interference management in D2D communication, HARQ appears to be a viable technique to increase the robustness.  
HARQ can be sent either directly (i.e., from D2D receiver to transmitter) or indirectly (i.e., from D2D receiver to the eNB, and from the eNB to D2D transmitter)~\cite{Lin2013ComMag}. The direct mode poses less overhead to the eNB in comparison to indirect mode. Moreover, benefits from the ACK/NACK messages arrive to the transmitter with shorter delay.

}

\subsection{Potential Future Work}
\label{ss:future_works}
Here we elaborate on some of the possible research directions and open problems in D2D communications in cellular networks. In the following we list some open problems based on different research methodologies.

{\bf Theoretical work.} As we mentioned earlier, the use of mathematical tools and optimization techniques in the state-of-the-art \off{is} \on{are} very limited. The current literature definitely lacks optimal mode selection techniques and  interference and power control mechanisms. The queue stability analysis using techniques such as stochastic Lyapunov optimization 
can be also an interesting issue to tackle. This can be further extended to provide throughput-based utility, throughput-power tradeoff, delay bounds, and delay analysis of D2D communications in cellular networks. 

{\bf Architecture.} There is very little work explaining the required architecture in order to support D2D communications in cellular networks~\cite{doppler_device_2009,yang_solving_2013}. It is interesting to further investigate on  the capability of the current centralized cellular architecture to handle D2D procedures such as device discovery, D2D connection setup, cellular network registration process, interference control, resource allocation, security, and so on. \on{Similarly, software defined networking-oriented architectures soon will have to include D2D in the equation. Indeed, D2D needs to be studied in the more complex context of HetNets due to growing market interest for availability of multiple radio technologies deployed on mobile devices.}

{\bf Application.} A decade ago, D2D was first proposed for relaying purposes in cellular networks. To date, researchers proposed new use-cases for D2D communications in cellular networks such as multicasting~\cite{du_compressed_2012,zhou_intracluster_2013}, peer-to-peer communication~\cite{lei_operator_2012}, video dissemination~\cite{golrezaei_base_2012,li_device_2012,doppler_device_2009}, M2M communication~\cite{pratas_low_2013}, \on{and} cellular offloading~\cite{bao_dataspotting_2010}. We believe D2D communication can have more applications in the telecommunication world. For example, it would be interesting to see the application of D2D communication in social networking, location-aware services, vehicular networks, smart grids \cite{Niyato2011,Fey2012}, etc. 

{\bf Performance analysis.} As seen in Table~\ref{tb:eval}, the majority of the available literature is based on numerical or home grown simulations. Although these types of evaluation method are suitable for studying the potential gains, they are still far from reality due to simplified assumptions. We believe a performance evaluation using the existing network simulators such as NS3~\cite{ns3}, OPNET~\cite{opnet}, Omnet++~\cite{varga2007omnet++} or an experimental evaluation can help in revealing both real performance and new challenges of D2D communications in cellular networks.
\section{Conclusion}
\label{s:conclusion}

In this work, we provided an extensive survey on the available literature on D2D communications in cellular networks. We categorized the available literature based on the communication spectrum of D2D transmission into two major groups, namely, inband and outband. The works under inband D2D were further divided into underlay and overlay. Outband D2D related literature \off{were} \on{was} also sub-categorized as controlled and autonomous. 

The major issue faced in underlay D2D communication is the power control and interference management between D2D and cellular users. Overlay D2D communication does not have the interference issue because D2D and cellular resources do not overlap.  However, this approach allocates dedicated cellular resources to D2D users and has lower spectral efficiency than underlay. In outband D2D, there is no interference and power control issue between D2D and cellular users. Nevertheless, the interference level of the unlicensed spectrum is uncontrollable, hence, QoS guaranteeing in highly saturated wireless areas is a challenging task.  

We also discussed the weaknesses and strength of the existing literature. We pointed out the shortcomings of current works and proposed potential future research directions.  Our survey showed that D2D communication in cellular networks is immature and there are still numerous open issues \on{such as interference management, power control, etc}. We also shed light on some possible research directions \off{in terms of analysis and evaluation} \on{needed to improve the understanding of D2D potentialities for real world applications}.

\bibliographystyle{IEEEtran}
\bibliography{biblio}

\end{document}